\documentclass[iop,numberedappendix]{emulateapj}

\usepackage{url}
\usepackage{enumerate}
\usepackage{graphics}
\usepackage{amsmath}
\usepackage{graphicx}
\usepackage{epstopdf}
\usepackage{color} 
\usepackage{rotating}
\usepackage{lipsum}
\usepackage{tikz}
\newcommand{\cfbox}[2]{    \colorlet{currentcolor}{.}    {\color{#1}    \fbox{\color{currentcolor}#2}}}

\usepackage[normalem]{ulem} 
\usepackage[flushleft]{threeparttable}

\newcommand{\kms}{\mbox{\,km$\,$s$^{-1}$}}

\newcommand{\msun}{\mbox{$\rm M_{\sun}$}}

\newcommand{\hii}{H \scriptsize{II}\normalsize\,}

\usetikzlibrary{plotmarks}

\usepackage[backref,breaklinks,colorlinks,citecolor=blue]{hyperref}
\usepackage[all]{hypcap}

\def\nlq{}

\shorttitle{\it Galactic Mini-starburst complexes} 
\shortauthors{Quang Nguyen Luong et al.}

\begin{document}

\title{The scaling relations and star formation laws of mini-starburst complexes}
\author{Quang Nguy$\tilde{\hat{\rm e}}$n-Lu{\hskip-0.65mm\small'{}\hskip-0.5mm}o{\hskip-0.65mm\small'{}\hskip-0.5mm}ng\altaffilmark{1,2,3,\dag}, 
Hans V. V. Nguy$\tilde{\hat{\rm e}}$n\altaffilmark{4}, 
Fred\'erique Motte\altaffilmark{5},				
Nicola Schneider\altaffilmark{6}, 	
Michiko Fujii\altaffilmark{7,8}, 			                                                        
Fabien Louvet\altaffilmark{9}, 
Tracey Hill\altaffilmark{10}, 
Patricio Sanhueza\altaffilmark{1}, 
James O. Chibueze\altaffilmark{11}, 		
and Pierre Didelon\altaffilmark{12} 		
}

\altaffiltext{1}{NAOJ Chile Observatory, National Astronomical Observatory of Japan, 2-21-1 Osawa, Mitaka, Tokyo 181-8588, Japan}
\altaffiltext{2}{Korea Astronomy and Space Science Institute, 776 Daedeok daero, Yuseoung, Daejeon 34055, Republic of Korea}
\altaffiltext{3}{CITA, University of Toronto, 60 St. George Street, Toronto, ON M5S~3H8, Canada}
\altaffiltext{4}{Max-Planck-Institut f\"ur Radioastronomie, Auf dem H\"ugel 69, D-53121 Bonn, Germany}
\altaffiltext{5}{Institut de Plan\'etologie et d'Astrophysique de Grenoble, LAOG (UMR 5571), Université de Grenoble, BP 53, F-38041 Grenoble Cedex 09, France}
\altaffiltext{6}{I. Physik. Institut, University of Cologne, 50937 Cologne, Germany}
\altaffiltext{7}{Theoretical Astronomy Division, National Astronomical Observatory of Japan, 2-21-1 Osawa, Mitaka, Tokyo 181-8588, Japan}
\altaffiltext{8}{Department of Astronomy, The University of Tokyo, Hongo 7-3-1, Bunkyo-ku, Tokyo, 113-0033, Japan}
\altaffiltext{9}{Departamento de Astronom\'ia, Universidad de Chile, Santiago, Chile}
\altaffiltext{10}{Joint ALMA Observatory, 3107 Alonso de Cordova, Vitacura, Santiago, Chile}
\altaffiltext{11}{Department of Physics and Astronomy, Faculty of Physical Sciences, University of Nigeria, Carver Building, 1 University Road, Nsukka, Nigeria}
\altaffiltext{12}{Laboratoire AIM Paris-Saclay, CEA/IRFU - CNRS/INSU - Universit\'e Paris Diderot, Service d'Astrophysique, B\^at. 709, CEA-Saclay, F-91191, Gif-sur-Yvette Cedex, France}
\altaffiltext{\dag}{EACOA Fellow at NAOJ, Japan \& KASI, Korea, \url{ quang.nguyen-luong@nao.ac.jp,quangnguyenluong@kasi.re.kr}}

\begin{abstract}
The scaling relations and the star formation laws for molecular cloud complexes in the Milky Way is investigated using data from the $^{12}$CO 1--0 CfA survey and from the literatures. We compare their masses $M_{\rm gas}$, mass surface densities $\Sigma_{M_{\rm gas}}$, radii $R$, velocity dispersions $\sigma$, star formation rates $SFR$, and SFR densities $\Sigma_{\rm SFR}$ with those of structures ranging from cores, clumps, Giant Molecular Clouds (GMCs), to Molecular Cloud Complexes (MCCs), and to Galaxies, spanning 8 orders of magnitudes in size and 13 orders of magnitudes in mass. MCC are mostly large ($R>50$ pc), massive ($\sim 10^{6}$\,\msun) gravitationally unbound cloud structures. This results in the following universal relations:

\begin{center}
$\sigma\sim R^{0.5}$, $M_{\rm gas}\sim R^{2}$, $\Sigma_{\rm SFR}\sim \Sigma_{M_{\rm gas}}^{1.5}$, ${SFR}\sim {M_{\rm gas}}^{0.9}$, and ${SFR}\sim {\sigma}^{2.7}$.
\end{center}

Variations in the slopes and the coefficients of these relations are found at individual scales signifying different physics acting at different scales. Additionally, there are breaks at the MCC scale in the $\sigma-R$ relation and between the starburst and the normal star-forming objects in the $SFR-M_{\rm gas}$ and $\Sigma_{\rm SFR}$-$\Sigma_{\rm M_{\rm gas}}$ relations. Therefore, we propose to use the Schmidt-Kennicutt diagram to distinguish the starburst from the normal star-forming structures by applying a $\Sigma_{M_{\rm gas}}$ threshold of $\sim100$\,\msun pc$^{-2}$ and a $\Sigma_{\rm SFR}$ threshold of 1\,\msun yr$^{-1}$ kpc$^{-2}$. 
Mini-starburst complexes are gravitationally unbound MCCs that have enhanced $\Sigma_{\rm SFR}$ ($>$1\,\msun yr$^{-1}$ kpc$^{-2}$), probably caused by dynamic events such as radiation pressure, colliding flows, or spiral arm gravitational instability.
Because of the dynamical evolution, gravitational boundedness does not play a significant role in characterizing the star formation activity of MCCs, especially the mini-starburst complexes, which leads to the conclusion that the formation of massive stars and clusters is dynamic. {\nlq We emphasize the importance of understanding mini-starburst in investigating the physics of starburst galaxies.}
\end{abstract}

\keywords{
stars: formation, ISM: clouds, ISM: structure, (ISM:) evolution, methods: observational, Galaxy: evolution\\
{\it (Submitted on May 02, 2016; Accepted to publish on ApJ on Oct 09, 2016)}}

\section{Introduction}
Molecular gas is an indispensable element of the galactic ecological system and exists as coherent cloudy structures having different sizes: core ($<$0.1 pc), clump (0.1--1 pc),  molecular cloud (GMC, 1--10 pc), and molecular cloud complex (MCC, 10--100 pc) \citep{blitz99}. They are the stellar nursery, therefore, large surveys of molecular clouds in the Milky Way are necessary to understand their global star formation activities and connection to the global properties of the Galaxy.

The birth of the millimeter-wave radio astronomy, and subsequently, of the observations of Carbon Monoxide (CO) emission opened a new window into the molecular gas \citep{wilson70}. Since then, molecular gas was discovered progressively in star forming regions \citep{lada74}, in the Galaxy's diffuse interstellar medium \citep{cohen80}, and in other galaxies \citep{elmegreen80}, thus revealed the ubiquity of molecular gas.

A milestone of the wide-field molecular gas observation is the CO 1--0 almost-all-sky survey performed by the CfA 1.2 m telescopes \citep{dame86,dame01}. This survey formed a coherent and high spectral resolution basis for subsequent follow-up surveys.
Since then, surveys of different excitation transitions or isotopologues were performed with larger telescopes to reach higher angular resolutions, for example: CO 1--0 from the Three-mm Ultimate Mopra Milky Way Survey (THRUMMS, \citealt{barnes15}), CO 3--2 from the CO 3--2 High-resolution Survey of the Galactic Plane (COHRS, \citealt{dempsey13}), or the $^{13}$CO 1--0 from the Galactic Ring Survey (GRS, \citealt{jackson06}).
However, the CfA survey is still pertinent to
create a large catalog of GMCs and MCCs, and to study the mutual scaling relations between different physical properties such as mass, size, velocity dispersion, and star formation rate (SFR).   
 
The existence of the universal relations between different physical properties of star-forming structures is a suggestion that the basic physics governing different objects are similar and only scale with their sizes. Confirmation or refutation of this argument needs an investigation of these relations on datasets that cover the entire physical scale range.  
Early on, \cite{larson81} pioneered in deriving the universal scaling relations between mass $M$, radius $R$, and velocity dispersion $\sigma$ of molecular clouds; later, they are named as the Larson's relations. First, the linewidth-size relation $\sigma\propto R^\beta$ with $\beta=0.38$ describes the structure of molecular clouds as fragmentation due to Kolmogorow-like turbulent cascade. Second, a linear correlation between the virial mass and total mass predicts the virial equilibrium state of GMCs. Third, the inverse relationship between the mean density and size implies a mass-size relation $M\propto R^\alpha$ with $\alpha=2$.

However, there are arguments against the universal scaling relations. 
Fore instance, \cite{lombardi10} and \cite{kauffmann10} derived a $M\propto R^\alpha$ relation with $\alpha\sim1.2-1.6$ for substructures inside individual GMC. \cite{heyer09} suggested that the coefficient of the velocity dispersion-radius relation for Galaxy's GMCs scales with the surface density. \cite{hughes13} found no trivial scaling relations between the three quantities: mass, size and velocity dispersion for GMCs and MCCs in M51, M33, and the Large Magellanic Cloud. \cite{utomo15} and  \cite{rebolledo15} found similar results for GMCs with average radii of 20 pc in NGC 4526 and MCCs with averaged radii of $\sim140-180$ pc in nearby galaxies, respectively. Furthermore, \cite{schneider04} showed that the coefficients of these relations strongly depend on the cloud structure identification algorithm that was used.

Likewise, \cite{schmidt59} and \cite{kennicutt98} found a correlation between gas surface density, $\sigma$, and Star Formation Rate density, $\Sigma_{\rm SFR}$, in the form of ${\rm \Sigma_{SFR}}  = A_{\rm KS}\times {\rm \sigma}^N$. 
Galaxy types, spatial resolutions, SFR tracers, and gas tracers are among the multiple factors that can change the power law index $N$ and the normalization factor $A_{\rm KS}$ \citep{deharveng94,gao04,bigiel08,daddi10}. 

The recent combinations of observations of Galactic clouds of $\sim$1--10~pc and galaxy populations of $\sim$1--10~kpc revealed a large spread in the $\Sigma_{SFR}-\sigma$ diagram \citep{heiderman10,evans14,willis15}. For example, the $\Sigma_{SFR}-\sigma$ relation of low-mass Galactic clouds derived by  \cite{heiderman10} has a steeper slope than the extragalactic one derived by \cite{kennicutt98}. Power law indexes within the massive star forming complexes were found to scatter from 1.7 to 2.8 \citep{willis15}. Various modifications of the Schmidt-Kennicutt relation were proposed: normalizing $\sigma$ by dynamical  \citep{daddi10} or freefall timescales \citep{krumholz12}, considering the SFR-dense gas relation as $\Sigma_{\rm SFR}\propto \Sigma_{\rm dense~gas}$ \citep{lada12,evans14}, or replacing the surface density by volume density quantities to produce a linear relation between star formation rate volume-density, $\rho_{SFR}$, and mass volume-density, $\rho_{\rm gas}$ \citep{evans14}.

Studies trying to link these scaling relations across different spatial scales, from local Galactic clouds to global galaxy, seem to miss the MCC population, also called Giant Molecular Associations, having sizes and masses of $>$ 50~pc and $\sim$ {$10^6-10^7~\msun$}, respectively \citep{nguyenluong11}. MCCs are important as they are the largest cloud agglomeration in a galaxy and massive star formation is linked with a special category of MCCs, the mini-starburst complexes
\citep[e.g.,][]{motte03,louvet14}. 
They are the birthplaces of massive OB stars and Young Massive Clusters (YMCs), thus are important in maintaining the chemical, energy, and mass balance of hosting galaxies \citep{bressert12}. This is especially true for starburst galaxies such as the Antennae
merger system, which undergoes starburst events and contains many YMCs and mini-starburst MCCs \citep{herrera12,whitmore14}.

Motivated by the needs of quantifying the scaling relations between different physical quantities of MCCs in the Milky Way, we use the CfA CO survey to catalog and characterize the physical properties of our MCCs and use radio continuum data to measure their SFR. 
We discuss the data and the method of identifying sources in Sections~\ref{sect:data} and \ref{sect:souID}. In Section~\ref{sect:MCCproper}, we derive the physical properties and SFR (density) of MCCs.  Section~\ref{sect:larson} and Section~\ref{sect:sf} will examine the scaling relations between different cloud properties and the star formation laws for the all cloud structures ranging from GMCs to Galaxies. We elaborate more the division of (mini)starburst and normal-star forming objects by using the Schmidt-Kennicutt diagram and the possible sequence of forming mini-starburst in Section~\ref{sect:discussion}.

\section{Data}
\label{sect:data}

\subsection{$^{12}$CO Data from the CfA Survey}
\begin{figure*}[htbp]
\centering
$\begin{array}{c}
\hskip -1cm \includegraphics[angle=0,width=18.cm]{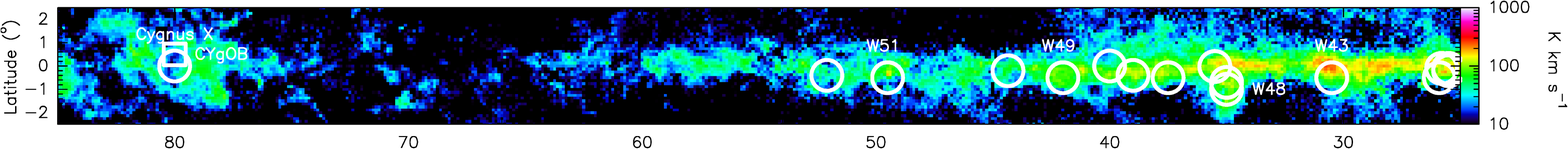} \\
\hskip -1cm \includegraphics[angle=0,width=18.cm]{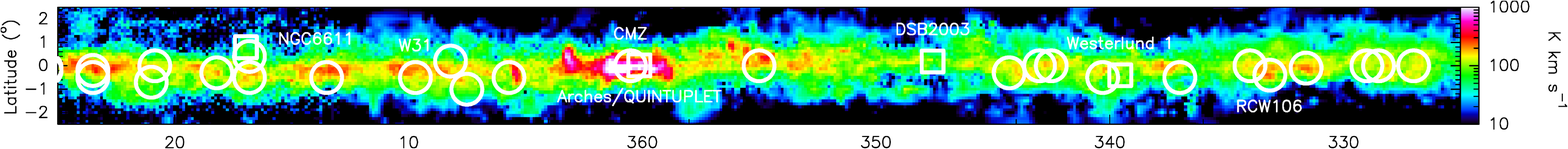} \\
\hskip -1.cm \includegraphics[angle=0,width=18.cm]{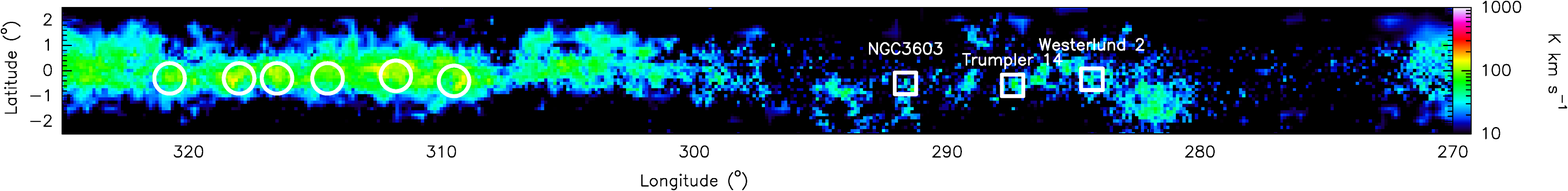} 
\end{array}$
\caption{The integrated $^{12}$CO~1--0 map of the CO CfA all sky survey, over which mini-starburst MCCs are indicated by circles. Names of well-known MCCs are labelled. Young Massive Clusters and OB Associations from \cite{portegies-zwart10} are marked as squares, respectively.}
\label{fig:MScanGP}
\end{figure*}

The $^{12}$CO~1--0 data from the CO all-sky survey is used to catalog and characterize MCCs in the Galaxy. This survey was made with the CfA 1.2~m telescopes located in both northern and southern hemispheres. The observations are sub-Nyquist sampled with an effective angular resolution of 8\arcmin.8. The spectral cube contains data obtained from various independent observations starting in 1986 \citep{dame86,bronfman89} and ending in 2001 \citep{dame01}, stored in the CO survey archive\footnote{ http://www.cfa.harvard.edu/rtdc/CO/}. The combined data cube is provided in main beam antenna temperature and has a sensitivity of $\sim$ 1.5 K per 0.65 \kms\,, although individual surveys have sensitivity $\sim$ 0.12--1  K  per 0.65-\kms channel. We use the \lq\lq whole galaxy cube" covering the entire 360\degr\, longitude range and -40\degr\, to 40\degr\, latitude range.  
The integrated map of the Galactic Plane, where MCCs reside, is shown in Figure~\ref{fig:MScanGP}.

\subsection{Radio continuum}
To derive the SFRs of MCCs, we use the 21 cm radio continuum data from the VLA Galactic Plane Survey (VGPS, \citealt{stil06}), the Canadian Galactic Plane Survey (CGPS, \citealt{taylor03}), and the Southern Galactic Plane Survey (SGPS, \citealt{haverkorn06}). VGPS has a FWHM beam of 1\arcmin\, and covers the Galactic longitude 18--67\degr, CGPS  has a FWHM beam of 1\arcmin\, and covers the Galactic longitude 63--175\degr, and SGPS has a FWHM of 2.2\arcmin\, covering the Galactic longitude 253--358\degr.

\subsection{Complementary data}
\label{sect:comdata}
We complement our data with literature data to increase the dynamical ranges of all parameters. For the Larson relations (Section\,\ref{sect:larson}), we add cloud properties data from the following sources: 
\begin{itemize}
\item clumps/GMCs:  \cite{maruta10},  \cite{onishi02},  \cite{shimajiri15},  \cite{heyer09}, \cite{roman-duval10},   \cite{evans14}.
\item MCCs: \cite{garcia14}, \cite{murray11},     \cite{donavan-meyer13}, \cite{rosolowsky07},   \cite{miura12,miura14}, \cite{wei12}.
\item Galaxies: \cite{leroy13},   \cite{tacconi13},  \cite{genzel10}.
\end{itemize}

For the Schmidt-Kennicutt scaling relation  (Section\,\ref{sect:sf}), we add SFR data from the following sources: 
\begin{itemize}
\item GMCs:  \cite{heiderman10}, \cite{lada10}, \cite{evans14}.
\item MCCs: \cite{garcia14}, \cite{bolatto08}, \cite{murray11}, \cite{donavan-meyer13}, \cite{rosolowsky07}, \cite{miura12,miura14},  \cite{wei12}.
\item Galaxies: \cite{leroy13},  \cite{tacconi13},  \cite{genzel10}.
\end{itemize}

For the galaxies sample, we calculate the velocity dispersion assuming that the entire galaxy is a dynamical system, thus we do not distinguish between elliptical and disk galaxies or starburst and normal galaxies. We convert the orbital time from \cite{leroy13} sample or the dynamical time from \cite{genzel10} and \cite{tacconi13} samples to a rotation velocity as $V_{\rm rot} = \frac{2\pi R}{t_{\rm dyn,orb}}$. 
Then, the velocity dispersion is written as
$\sigma = \frac{2\pi^2 GR\sigma}{1.5V_{\rm rot}}$. The radi used here are either the optical B-band 25th magnitude isophote \citep{leroy13} or the half-light radii \citep{genzel10,tacconi13}.
The local galaxies in \cite{leroy13} have velocity dispersions smaller than their rotation velocity, but the high-z galaxies in \cite{genzel10,tacconi13} have velocity dispersion larger than rotation velocity. This is probably, because the high-z galaxies are highly turbulent than the local galaxies.

Beside that, we also add data from individual mini-starburst regions in a galaxy at $z = 1.987$ \citep{zanella15}, in SDP81 galaxy at $z = 3.042$ \citep{hatsukade15}, and in Arp 220  \citep{scoville13}. 

Systematic errors caused by different measurements are unavoidable but we assume that these errors have little affect on our analysis. For example, errors caused by mass determinations from different tracers such as $^{12}$CO, $^{13}$CO, or dust emission; or errors caused by the variation of the abundances for the CO/H$_{2}$ conversion as a function of position in the galaxy; or errors caused by the exclusion of the CO-dark gas, are not taken into account. So, the mass estimates are the lower limits of the true values. 
We are aware of these error sources, but due to our large statistical sample, and assuming an uncertainty of typically 50\%, we are still able to derive robust scaling relations. 
Actually, these errors can be safely ignored because they typically vary around about 30\%-100\% of the measured values and all relations that we will discuss are presented in log-log space.

\section{Source identification and Distances}
\label{sect:souID}
Initially, we use the Duchamp\footnote{http://www.atnf.csiro.au/people/Matthew.Whiting/Duchamp} algorithm to decompose the 3D spectral cube. This program is designed to extract sources in large H {\scriptsize I} surveys and is the main extraction tool of the Australian Square Kilometre Array Pathfinder \citep{whiting12b}. Although it is optimized for H {\scriptsize I} source extraction, its utility towards molecular clouds and maser extraction has been proven \citep{carlhoff13,walsh15}. For our extraction, Duchamp detected successfully GMCs outside and inside the Galactic Plane. However, in the Galactic Plane, Duchamp tends to either break down the emission into too many individual clouds if we set a weak merging condition or merge to too large structures. This is likely the same problem occured for other detection alogrithms (Gaussclumps, \citealt{stutzki89}; Clumpfind, \citealt{williams94}; Dendogram, \citealt{rosolowsky08} ). Finally, we decided for a conservative approach to select MCCs by eye inspection.

We identify strong CO peaks in the integrated intensity and position-velocity maps of \cite{dame01} and then average CO spectra over $0.5\degr\times0.5\degr$ areas surrounding the CO peaks that correspond to $\sim$13--52~pc boxes at 1.5--6~kpc distances. The global velocity extent measured from these first-guess integrated spectra are used to integrate CO lines and create first-guess integrated maps of the MCC. For each MCC, we then iteratively refine its associated area, $A$, and velocity extent until they are properly distinguished both in velocity and space from a background of three times the local rms. 
By setting the local rms as threshold, we can trace all material down to the observational sensitivity level, but not having homogeneous column density threshold across all MCCs. 
We fit Gaussian profiles to the average spectrum of each MCC to derive its velocity dispersion, $\sigma_{\rm CO}$, and its velocity-integrated intensity, $W_{\rm CO}$.
{\nlq To summarize, the main criteria to include a cloud in MCC are that their velocity range is not more than 15\,\kms\, far from the bulk velocity of the cloud and it is connected to the main and other clouds by diffuse gas features.} This method was proven to be more suited to identify MCCs than automatic detection, as in the cases of the W43 \citep{nguyenluong11b} and the RCW 106 \citep{nguyenhan15}.

After identifying a sample of 44 MCCs (Table\,\ref{tab:MCCproperties}), we estimate their distances to the Sun following this sequential scheme: (1) assigning a parallax distance if it is available from the Bar and Spiral Structure Legacy (BeSSeL,  \citealt{reid14})\footnote{http://bessel.vlbi-astrometry.org} and the VLBI Exploration of Radio Astrometry (VERA, \citealt{honma07})\footnote{http://veraserver.mtk.nao.ac.jp} projects or other individual parallax measurements from the literature; (2) assigning a photometric distance if it is available from the literature; (3) assigning an averaged ambiguity-resolved kinematic distance if it is available from the literature; (4) calculating our own kinematic distance as 
$d_{\rm kin} = R_{\rm 0} {\rm cos}\left(l\right)\pm\sqrt{r^{2}-R_{\rm 0}^{2}{\rm sin}^{2}\left(l\right)}$ where 
$R_{\rm 0}$ is the Galactocentric radius of the Sun, $V_{\rm 0}$ is the orbital velocity of the Sun around the Galactic center, $V\left(r\right)$ is the rotation curve and  $V_{\rm r}$ is the radial velocity of the cloud. {\nlq The result is 11 out of 44 have parallax distances, 1 has photometric distance, 8 have kinematic distances with ambiguity resolved by H{\scriptsize I} absorption, and 24 have near kinematic distances. }

\section{Physical properties}
\label{sect:MCCproper}

\begin{table*}[htpb]
 \small  \label{tab:MCCproperties}
 \setlength{\tabcolsep}{0.07cm}
 \renewcommand{\arraystretch}{0.75}
 \begin{tabular}{lrrccccccccccccccccc}
  \hline
  \hline
  \multicolumn{1}{c}{Complex} &	 \multicolumn{1}{c}{l} &	 \multicolumn{1}{c}{b} &	 \multicolumn{1}{c}{$V_{\rm LSR}$} &	 \multicolumn{1}{c}{$\sigma$} &	 \multicolumn{1}{c}{$d$} 
  &	 \multicolumn{1}{c}{$A$}&	 \multicolumn{1}{c}{$R$} &	 \multicolumn{1}{c}{$L_{\rm CO}$} &	 \multicolumn{1}{c}{$M_{\rm gas}$} &	 \multicolumn{1}{c}{$\Sigma_{\rm gas}$} 
  &	   \multicolumn{1}{c}{$S_{\rm 21 cm}^{\rm int}$} & \multicolumn{1}{c}{SFR} &	 \multicolumn{1}{c}{$\Sigma_{\rm SFR}$}
  &	 	 \multicolumn{1}{c}{$\alpha_{\rm vir}$} \\ 
  
  \multicolumn{1}{c}{name} &	 \multicolumn{1}{c}{} &	 \multicolumn{1}{c}{} &	 \multicolumn{1}{c}{} &	 \multicolumn{1}{c}{}
  &	 \multicolumn{1}{c}{} &	 \multicolumn{1}{c}{$\times 10^{-2}$}&	 \multicolumn{1}{c}{} &	 \multicolumn{1}{c}{$\times 10^{3}$} 
  &	 \multicolumn{1}{c}{$\times 10^6$} &	 \multicolumn{1}{c}{} &	 \multicolumn{1}{c}{} &	
  \multicolumn{1}{c}{} & \multicolumn{1}{c}{} 	 	& \multicolumn{1}{c}{} \text
  {}  \\ 
  
  \multicolumn{1}{c}{} &	 \multicolumn{1}{c}{(\degr)} &	 \multicolumn{1}{c}{(\degr)} &		\multicolumn{1}{c}{(\kms)}  &		\multicolumn{1}{c}{(\kms)}
  &	 \multicolumn{1}{c}{(kpc)} &	 \multicolumn{1}{c}{(kpc$^{2}$)}&	 \multicolumn{1}{c}{(pc)} &	 \multicolumn{1}{c}{(L$_{\odot}$)} 
  &	 \multicolumn{1}{c}{(M$_{\odot}$)} &	 \multicolumn{1}{c}{(M$_{\odot}$\,pc$^{-2}$)}
  &	 \multicolumn{1}{c}{(Jy)} &	 \multicolumn{1}{c}{(M$_{\odot}$\,yr$^{-1}$)} &	 
  \multicolumn{1}{c}{\tiny{(M$_{\odot}$\,yr$^{-1}$kpc$^{-2}$})} &	  \multicolumn{1}{c}{} 	\\ 
 \hline 
\hline 
 G111 	        & 111.0	&-1.0	&-49	&7.9	&\cfbox{red}{3.34}	&5.7	&134	&7.2	&2.2	&38.7		&1860  &  0.014 	&0.25	&4.4		\\ 
\bf G80-CygnusX & 80.0	&0.0	&2	&6.6	&\cfbox{red}{1.5}	&3.7	&108	&40.7	&2.2	&59.9		&45759  &  0.062  &1.68	&2.5	\\ 
 
   G52...... 	 & 52.1	&-0.4	&55	&6.9	&\cfbox{green}{6.9}	&2.0	&79	&1.8	&1.9	&95.6		&2818  &  0.0725	&3.6	&2.3		\\ 
 \bf G49-W51....& 49.5	&-0.5	&57	&9.8	&\cfbox{red}{5.41}	&1.8	&76	&3.6	&2.5	&139.9		&10237  &  0.184 & 10.2	2& 3.4		 \\ 
  G44.4...... 	 & 44.4	&-0.2	&61	&7.2	&\cfbox{green}{9.3}	&1.8	&75	&2.7	&1.5	&82.7		&5272  &  0.070  &3.89	&3.1		\\   
 \bf G42-W49....& 42.0	&-0.5	&63	&9.5	&\cfbox{red}{11.1}	&2.8	&93	&1.5	&4.9	&179.2		&6069  &  0.51  &17.86	&2.0		 \\  
 G40........ 	 & 40.0	&0.0	&32	&6.1	&2.2	&3.4	&104	&13.0	&1.8	&52.1		&7783  &  0.026   	&0.77	&2.5		 \\  
 G39........ 	 & 39.0	&-0.4	&65	&27.3	&\cfbox{green}{12.1}	&13.1	&204	&33.1	&15.7	&120.2		&6447  &  0.079  	&0.60	&11.3		 \\   
 \bf G35-W48....& 35.0	&-1.0	&46	&14.0	&\cfbox{red}{3.27}	&3.3	&101	&15.3	&4.5	&138.0	&8379  &  0.075 &2.27	&5.2		 \\  
 \bf G30-W43.... & 30.5	&-0.5	&93	&16.3	&\cfbox{red}{5.5}	&2.8	&94	&11.1	&9.3	&329.5		&15282  &  0.320  	&11.40	&3.1		 \\ 
 G25.9...... 	 & 25.9	&-0.5	&100	&10.0	&5.4	&2.5	&88	&4.5	&3.7	&148.5		&12688  &  0.255 	&10.24	&2.8	 \\  
 G25.5+52\kms 	 & 25.5	&-0.2	&52	&11.1	&3.4	&1.1	&58	&5.2	&1.7	&154.0		&15012  &  0.120  &10.91	&5.0	 \\  
 G25.5+102\kms 	 & 25.5	&-0.1	&102	&11.8	&5.5	&2.7	&93	&5.6	&4.6	&169.8		&15164  &  0.317 &11.75 	&3.2		 \\ 
G23.7+60\kms 	 & 23.7	&-0.5	&60	&9.9	&\cfbox{red}{6.21}	&2.0	&80	&4.4	&4.7	&230.9	&15342  &  0.407 	&20.40	&2.0	 \\ 
 G24+100\kms 	 & 23.5	&-0.2	&93	&21.1	&5.9	&3.2	&101	&13.9	&13.2	&411.9		&17171  &  0.413 	&12.91	&4.0		 \\  
  G21........ 	 & 21.0	&-0.7	&51	&13.2	&3.6	&1.2	&61	&7.3	&2.6	&216.8		&11690  &  0.1048 	&6.10	&4.8	 \\ 
 G20.9..... 	 & 20.9	&0.0	&32	&14.7	&2.6	&0.6	&44	&5.1	&1.0	&151.3	&12909  &  0.060  &10.0	0 &11.8		 \\ 
  G18.2...... 	 & 18.2	&-0.3	&47	&10.3	&3.6	&1.2	&61	&7.3	&2.6	&217.7		&5044  &  0.045 	&3.75	&3.0	 \\ 
  G16.8-M16/M17.... 	 & 16.8	&0.4	&23	&5.9	&\cfbox{red}{1.98}	&0.5	&39	&4.7	&0.7	&140.7		&8560  &  0.031 	&6.20	&2.3		 \\ 
  \bf G13.5-W33.. 	 & 13.5	&-0.5	&24	&36.1	&\cfbox{red}{2.92}	&1.2	&61	&29.8	&4.7	&393.2		&439  &  0.00175 	&0.149	&19.8		 \\ 
 \bf G10-W31.... 	 & 9.7	&-0.5	&22	&15.7	&\cfbox{red}{4.95}	&2.3	&85	&14.0	&9.5	&416.2		&-	&-	&1.0	&2.6		 \\ 
  G8.2....... 	 & 8.2	&0.2	&17	&8.4	&2.9	&0.8	&49	&6.3	&1.5	&188.4		&-	&-	&1.0	&2.8	 \\ 
 G7.5....... 	 & 7.5	&-1.0	&17	&8.7	&3.0	&0.8	&51	&7.9	&2.0	&234.4		&-	&-	&1.0	&2.3		 \\ 
 G3.5....... 	 & 5.7	&-0.5	&13	&8.6	&3.1	&2.8	&93	&19.0	&5.0	&179.7		&-	&-	&0.9	&1.6	 \\ 
 \bf G0-CMZ..... 	 & 0.5	&0.0	&-6	&23.1	&\cfbox{red}{7.9}	&51.6	&405	&74.8	&128.5	&249.1		&-	&-	&0.9	&2.0	\\ 
 G355....... 	 & 355.0	&0.0	&95	&20.0	&6.1	&0.8	&50	&2.7	&2.8	&344.2		&824  &  0.0212 	&2.50	&8.4		 \\ 
 G344....... 	 & 344.3	&-0.3	&-71	&8.1	&4.8	&2.6	&91	&3.3	&2.1	&80.4		&703  &  0.0112 	&0.423	&3.3 \\  
 G343....... 	 & 343.0	&0.0	&-28	&8.0	&2.7	&2.6	&91	&17.5	&3.4	&129.7		&426  &  0.00214 &0.077	&2.0		 \\  
 G342-127\kms 	 & 342.5	&0.0	&-127	&8.1	&\cfbox{green}{8.10}	&3.5	&105	&2.2	&2.5	&71.3		&245  &  0.007 	&0.17	&3.2		 \\ 
 G342-79\kms 	 & 342.5	&0.0	&-79	&7.8	&\cfbox{green}{4.74}	&2.3	&85	&3.4	&2.4	&102.8		&245  &  0.004 	&0.173	&2.5	 \\  
 G340....... 	 & 340.3	&-0.5	&-37	&14.1	&\cfbox{green}{3.41}	&0.8	&51	&10.4	&2.6	&309.4		&684  &  0.00426 	&0.50	&4.6		 \\  
 G337....... 	 & 337.0	&-0.5	&-118	&8.9	&\cfbox{green}{7.85}	&3.8	&109	&2.3	&2.9	&76.5		&2385  &  0.074 	&1.95	&3.5		 \\  
 G334....... 	 & 334.0	&0.0	&-87	&9.4	&4.9	&2.2	&84	&4.1	&2.7	&122.4		&2312  &  0.0384 	&1.77	&3.2	 \\  
 \bf G330-RCW106 	 & 333.1	&-0.4	&-46	&13.9	&\framebox{3.5}	&0.9	&53	&11.7	&3.2	&348.4		&3111  &  0.0278 	&3.22	&3.8	 \\  
 \bf G331-90\kms 	 & 331.6	&-0.1	&-93	&10.4	&\cfbox{green}{7.44}	&2.5	&88	&5.5	&4.0	&162.0		&1863  &  0.0348 &1.40	&2.8	 \\  
 G329-75\kms.. 	 & 328.5	&0.0	&-75	&30.2	&4.4	&4.1	&114	&28.0	&15.1	&369.5	&714  &  0.0095 	&0.244	&8.0	 \\  
 G329-25\kms..	 & 329.0	&0.0	&-45	&7.7	&3.0	&3.3	&103	&14.3	&3.5	&104.6		&917  &  0.00571 & 0.18	&2.0	\\  
 G327....... 	 & 327.0	&0.0	&-45	&7.8	&2.9	&1.8	&75	&11.3	&2.7	4&149.4		&2563  &  0.0149 	&0.83	&2.0	\\ 
 G320....... 	 & 320.8	&-0.3	&-62	&12.3	&4.0	&1.5	&69	&4.2	&1.9	&124.1		&811  &  0.00897 	&0.67	&6.5	 \\ 
 G318....... 	 & 318.0	&-0.3	&-44	&8.1	&3.0	&0.8	&51	&5.5	&1.4	&163.6		&833  &  0.00518 	&0.625	&2.9		 \\  
 G316.5..... 	 & 316.5	&-0.3	&-48	&9.1	&3.3	&1.0	&57	&6.1	&1.9	&180.0		&1224  &  0.0092 &0.1	&3.0	 \\ 
 G314....... 	 & 314.5	&-0.3	&-49	&8.9	&3.6	&1.2	&61	&7.4	&2.6	&219.3		&280  &  0.0025 &0.25	&2.1 \\ 
 G311....... 	 & 311.8	&-0.2	&-49	&8.9	&4.0	&2.4	&86	&11.7	&5.1	&214.4		&1099  &  0.0121  &0.5	&1.6	 \\ 
 G309....... 	 & 309.5	&-0.4	&-44	&11.2	&3.8	&1.8	&75	&9.0	&3.7	&205.9		&528  &  0.0053 	&0.28	&3.0		 \\ 

  \hline
  \hline
 \end{tabular}
 \hspace{2.5cm}
 \caption{   \small
     {\bf Cloud and SFR characteristics of the massive molecular cloud complexes (MCCs) in the Milky Way.}
  The distances quoted in red rectangles are parallax distances, in black rectangles are photometric distances, in green rectangles are ambiguity resolved kinematic distances, and the rests are near-kinematic distances. 
  References of parallax distances to several well-known MCCs: 
  G111 \citep{choi14}, 
  Cygnus X \citep{rygl12}, 
  W51 \citep{sato10}, 
  W49 \citep{zhang13},   
  G35-W48 \citep{zhang09}, 
  W43 \citep{zhang14}, 
  G23.7+60 \citep{sanna14}, 
  M16/M17 \citep{xu11}, 
  W33 \citep{immer13},
  W31 \citep{sanna14}, 
  CMZ \citep{reid09},  
  CMZ \citep{reid09},   
  NGC6334/NGC6357 \citep{chibueze14}.   For ambiguity resolved kinematic distances, we obtain them from \cite{jones12b} and \cite{anderson09}.
The SFRS of G10, G8.2, G7.5, G3.5, and CMZ MCCs are not calculated because the lack of radio continuum data toward the central molecular zone.}
 \end{table*}

\subsection{Cloud properties}
\label{sect:properties}

We use the CO 1--0 emission as a proxy to estimate the total gas mass content of molecular cloud structure. For this purpose, using the CO 1--0 alone is subjected to two major problems. First, CO may not trace all molecular hydrogen in molecular clouds and it miss the 'CO-dark' gas \citep{langer14}. Second, the CO 1--0 emission becomes optically thick quickly in the dense parts of the molecular clouds. Moreover, \cite{barnes15} proposed that the linear conversion from CO integrated intensity to H$_2$ gas column density might underestimate the true column density.
However, for the first-order global mass estimate, this is acceptable. The X factor $X=2\times10^{20}~{\rm cm}^{-2}$ ${\rm K}^{-1}\,(\rm \kms)^{-1}$ is used to convert the CO integrated intensity to H$_2$ gas column density as recommended by \cite{bolatto13}, which was established after an exhausted investigation of all possible measurements. 
The mean hydrogen molecular mass $m_{\rm H_2} = \mu m_{\rm H} = 2.8 m_{\rm H}$ which also accounts for helium contained in the gas is used to convert from number to mass column density. 
We note that additionally intrinsic uncertainties on the $X$ factor can also contribute to the error of the estimated column density and mass \cite[see, e.g.,][]{shetty11}.

Eventually, we derived the following physical parameters:

\begin{itemize}
 \item Velocity-integrated intensity $W_{\mbox{CO}}$ (K \kms) by integrating the channel maps within the velocity range derived from the Gaussian fit of the integrated spectrum,
 \item Equivalent radius $R = \sqrt{A/\pi}$ (pc) from the surface area $A$ (pc$^{2}$) measured in the integrated map,
 \item Velocity dispersion $\sigma = \frac{\Delta v_{\rm ,FWHM}}{\sqrt{\rm 8ln2}}$ (\kms) from the FWHM linewidth ${\Delta v_{\rm ,FWHM}}$ resulted from Gaussian fitting of the integrated spectra,
 \item CO luminosity $L_{\rm CO} = A  W_{\mbox{CO}} $ $\rm \left(K \kms pc^{2} \right) $,

 \item Total gas mass $M = L_{\rm CO}  X_{\rm CO}  m_{\mbox{\tiny H2}} \left(\msun\right)$,
 \item Gas surface density $\Sigma_{\rm gas} = \frac{M}{A}$ ${\rm \left( \msun pc^{-2}\right)}$,
 \item Virial parameter $\alpha_{\rm vir} = 5\sigma_{\rm 1D}^{2} R/GM$ with the gravitational constant G.
 
 \end{itemize} 

Our focus is on the 44 MCCs that are more massive than $10^6~\msun$ (Table~\ref{tab:MCCproperties}). All of them lie in the Galactic Plane within the longitude ranging from 0\degr to 90\degr\, or  310\degr to 355\degr\, and the latitude ranging from -1\degr to +1\degr, thus, mainly in the first and fourth quadrants (see Figure\,\ref{fig:MScanGP}).  Our MCCs coincide spatially with all massive GMCs having mass larger than $10^6~\msun$ in the other CO surveys \citep{heyer09,roman-duval10,garcia14}. 
All MCCs hosting massive star clusters characterized by \cite{murray11} are associated with
MCCs in our catalog. We also cover well-known MCCs such as W43 \citep{nguyenluong11b}, Cygnus~X \citep{schneider06}, W49 \citep{galvan-madrid13}, and W51 \citep{ginsburg15}. Therefore, we can conclude that our MCCs catalog is a robust catalog of nearby MCCs.
Table~\ref{tab:MCCproperties} lists, for each MCC the location (l, b, $V_{\rm VLSR}$), extent ($A$, $R$, $\sigma$), CO luminosity ($L_{\rm CO}$) along with its associated mass ($M_{\rm gas}$), gas surface density ($\Sigma_{\rm gas}$) and viral parameter ($\alpha_{\rm vir}$). 
Their typical sizes and masses range from 40~pc to 100~pc and from $1\times 10^6~\msun$ to $5\times 10^7~\msun$. One exception is the Central Molecular Zone which has a mass of  $1.3\times 10^8~\msun$ over a  area with equivalent radius of 406~pc.

\begin{figure}[hbtp]
\vskip -0cm
\centering
\includegraphics[angle=0,height=8.cm]{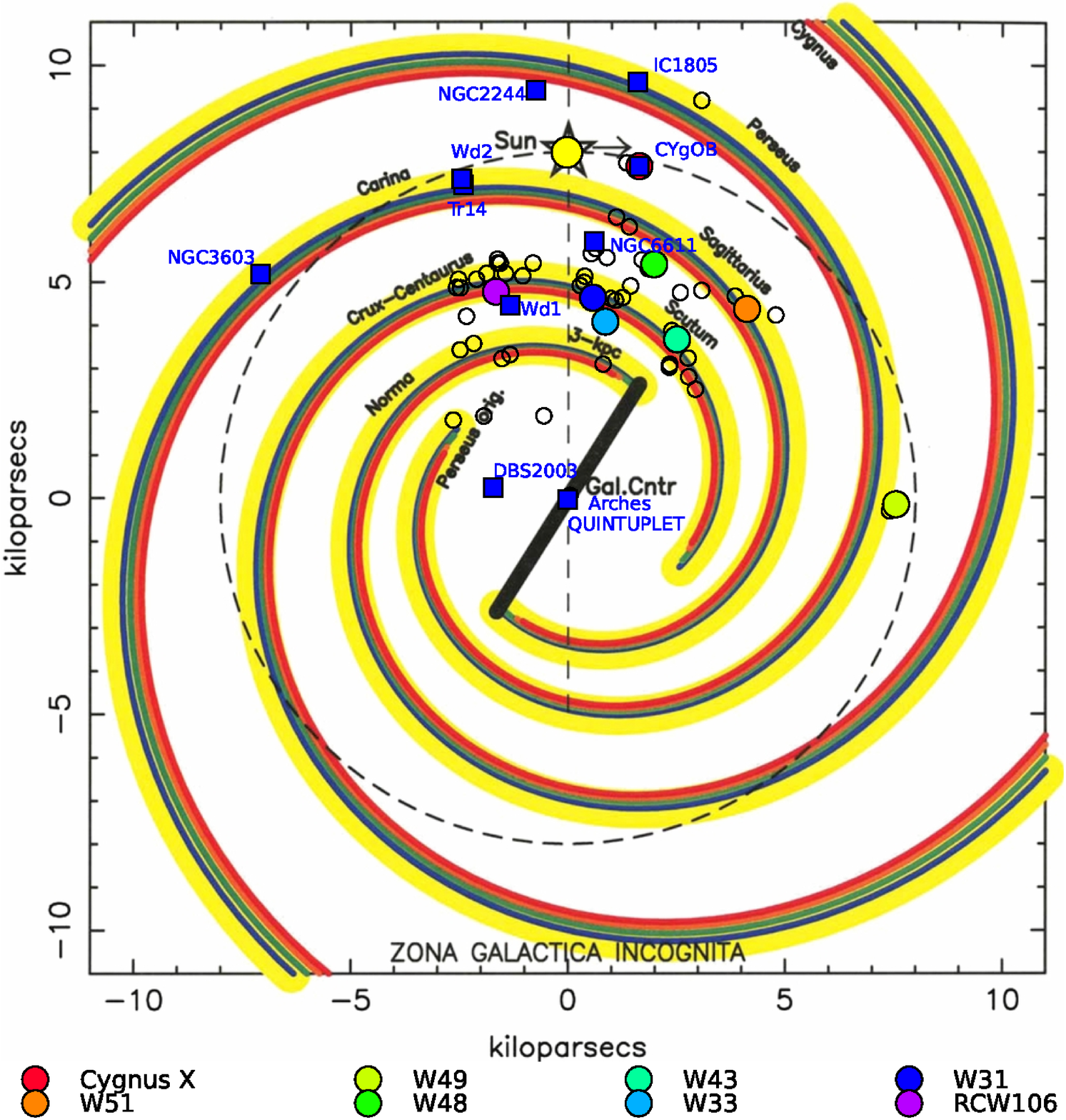}
\caption{Location of the mini-starburst MCCs over the spiral-arm structure of the Milky Way as modelled by \cite{vallee14}. Mini-starburst MCCs are indicated by circles. Well-known MCCs are labelled. Young Massive Clusters and OB Asscotiation from \cite{portegies-zwart10} are marked as squares, respectively.}

\label{fig:MSfaceon}
\end{figure}

To assess the detection completeness, we compare the number of MCCs in our catalog (44) with the total expected number of MCCs in the Milky Way from power-law mass distribution, $\frac{dN}{dm}\propto m_{\rm GMC}^{\gamma}$ with $\gamma=-1.5$ (e.g. \citealt{simon01}). Assuming that the Galactic molecular mass ranges from the minimum mass $M_{\rm L} = 10^2~\msun$ to the maximum mass $M_{\rm U} = 10^7~\msun$, we calculate the number of GMCs above a certain mass $m$ by integrating the mass distribution function over the mass range $[M_{\rm L}-m]$. This yields:
\begin{equation}
N(>m) = \frac{2-\gamma}{\gamma - 1} \times
\left( \frac{\left(\frac{M_{\rm U}}{m}\right)^{\gamma-1}-1}{1-\left(\frac{M_{\rm L}}{M_{\rm U}}\right)^{2-\gamma}} \right)
\times \frac{M_{\rm Tot}}{M_{\rm U}} \,\,\, ,
\end{equation}
where $M_{\rm Tot}=10^{9}~\msun$ is the total gas mass in the Milky Way \citep{dame01}.
A total number of $\sim$300 MCCs with masses larger than $10^6~\msun$ is expected in the entire Galaxy, about six times the number of detected MCCs. Our detection scheme mostly focuses on the near side of the Milky Way, covering only $\sim$1/5 of the Galactic plane, the part in the first and fourth quadrants (see Figure~\ref{fig:MSfaceon}) therefore the total number of 44 MCCs is complete within this Galactic region.

MCCs scatter along the Galactic plane and follow closely the spiral arms structure (see Figures~\ref{fig:MScanGP} and \ref{fig:MSfaceon}). Most (70\%) 
of the MCCs lie along the Scutum-Centaurus or Sagittarius arms, the two most prominent spiral arms in the Milky Way \citep{dame01}, and a few other lie along the Norma arm, Perseus arms, and interarm regions. We note that YMCs, the potential descendant of mini-starburst clouds, also lie along the spiral arms \citep{portegies-zwart10} (see also 
Figure~\ref{fig:MSfaceon}).
This is in line with extragalactic observations and numerical simulations, which find that the most massive, most turbulent, and most actively star-forming regions are located in the spiral arms rather than in the inter-arm zones \citep{koda12,dobbs06,fujimoto14}.
MCCs are concentrated to the midplane, they distribute within the latitude range $\pm1\degr$ or approximately $\sim$200~pc from the Galactic plane equator (see Figure~\ref{fig:MScanGP}).

\subsection{SFRs from radio continuum emission}
To calculate the immediate past (last $5\times 10^{6}$ yr, or the timescale of one OB star generation) SFR of MCCs, we follow the same approach as in \cite{nguyenhan15}. Following \cite{mezger67}, we use the 21 cm continuum emission to calculate the total Ly$\alpha$ continuum photons emitted by young massive stars that drive H{\scriptsize II} regions as

\begin{equation}
\frac{N_{\rm Ly\alpha}}{8.9\times10^{46}\,\text{s}^{-1}} = \frac{S_{\nu}}{\text{Jy}}   \left(\frac{\phantom{N} \nu \phantom{N}}{\text{GHz}}\right)^{0.1} \left(\frac{T_{e}}{10^{4}\,\text{K}}\right)^{-0.45}   \left(\frac{d}{\text{kpc}}\right)^{2} \, .
\label{SFR1}
\end{equation}
\noindent where $T_{e}=8000$~K \citep{wilson12} is the electron
temperature, $\nu=1.42$\,GHz is the observing frequency, and $d$ is
the distance to the region. 

Assuming that an O7 star with mass larger than 25\,\msun\, emits on average $N_{\rm Ly\alpha}={5\times10^{48}\,\text{s}^{-1}}$ \citep{martins05}, 
we calculate the SFR of the MCCs based on the SFR calibration for a full typical mass spectrum derived by \citep{murray10} as

\begin{equation}
\frac{{\rm SFR}}{\rm \msun yr^{-1}} = 4.1\times 10^{-54} \frac{ {N_{\rm Ly\alpha}}}{\rm s^{-1}} \, .
\end{equation}

The SFR is then calculated for the entire MCCs using the area derived from CO emission (see Table\,\ref{tab:MCCproperties}).
The SFR density of our MCCs range from 1 to 10  
$\rm ~\msun\, yr^{-1}\,kpc^{-2}$, which are in the high range of the Gould Belt dense cores \citep{heiderman10}, although the sizes of MCCs are hundreds times as large. SFR densities of MCCs are comparable with the SFR of  super giant H {\scriptsize II} regions in M33 \citep{miura14}.
The fact that these high SFR densities fill up the missing part in the SFR-Mass diagram derived by \cite{lada10} suggests that MCCs are good candidates to link the SFRs from local GMC scale to global galaxy scale (see Figure\,\ref{fig:ks2}).

When using the radio continuum flux to estimate the SFR of MCCs, a few assumptions have to be made, such as the mass spectrum over which the total stellar mass is calculated, the maximum cut-off mass of the mass spectrum, the non-dispersal property of the gas clouds, or the independence of the mass spectrum on the total gas mass. These assumption might add up uncertainties to our SFR measurements.

\section{Global Larson's scaling relations and break at MCC scales}
\label{sect:larson}

 \begin{figure}[hbtp]
\centering
$\begin{array}{ll}
\includegraphics[angle=0,width=9cm]{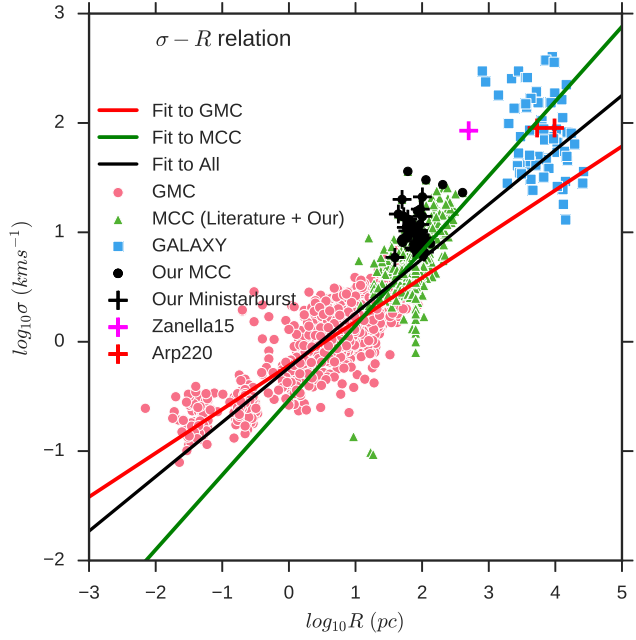} &
 \end{array}$
\vskip -0.5cm 
\caption{The Larson's scaling relation of velocity dispersion-radius.  All objects are divided into three categories: GMCs (size $<$ 10 pc), MCC (10 $<$ size $<$ 1000 pc), and Galaxy  (size $>$ 1000 pc). The continuous  lines are linear fits to the velocity dispersion-radius relations of GMC (pink), MCC (green), and all (black).  }  
\label{fig:larson1}
\end{figure}

 \begin{figure}[hbtp]
\centering
$\begin{array}{l}
\includegraphics[angle=0,width=9cm]{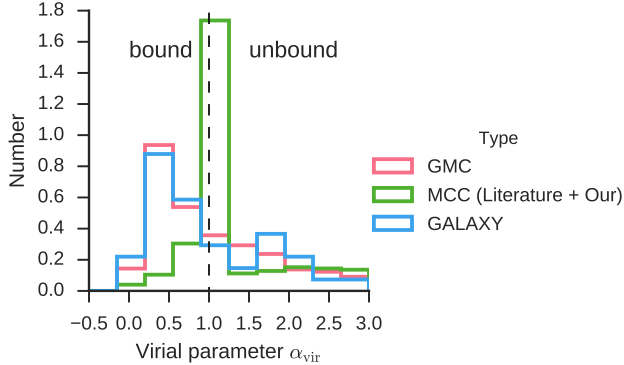} \\
 \end{array}$
\vskip -0.5cm 
\caption{The histogram of the virial parameters $\alpha_{\rm vir}$ of all objects which are divided into three categories: GMCs (size $<$ 10 pc), MCC (10 $<$ size $<$ 1000 pc), and Galaxy  (size $>$ 1000 pc). The dotted line divides the gravitaionally bound ($\alpha_{\rm vir} < 1$) and unbound ($\alpha_{\rm vir} > 1$) regimes.}  
\label{fig:larsonvirial}
\end{figure}

 \begin{figure}[hbtp]
\centering
$\begin{array}{ll}
\includegraphics[angle=0,width=9cm]{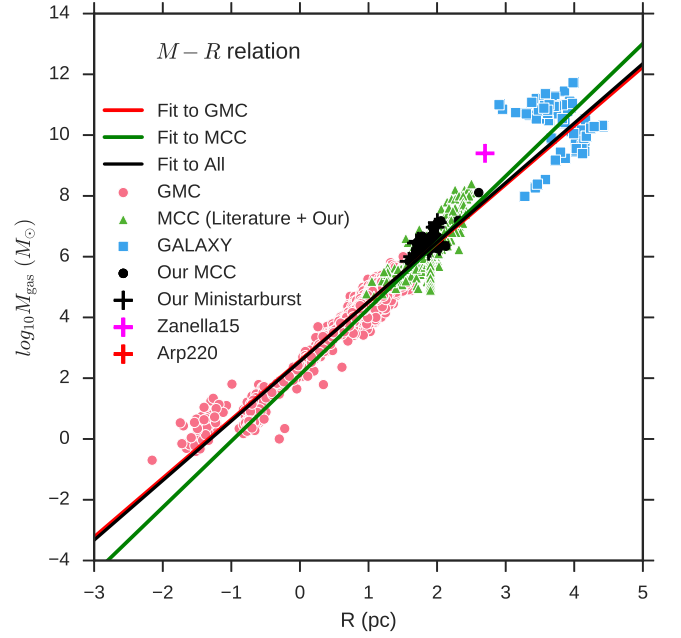} 
 \end{array}$
\caption{The Larson's scaling relation of mass-radius. Symbols are as in Figure~\ref{fig:larson1}. All objects are divided into three categories: GMCs (size $<$ 10 pc), MCC (10 $<$ size $<$ 1000 pc), and Galaxy  (size $>$ 1000 pc). The continuous  lines are linear fits to the mass-radius relations of GMC (pink), MCC (green), and all (black). }  
\label{fig:larson3}
\end{figure}

We analyse the scaling relations between different physical properties (mass, radius, gas mass density, and velocity dispersion) of cloud structures across 8 orders of magnitude in size and 13 orders of magnitude in mass using the most complete compilation of molecular cloud structures discussed in Section~\ref{sect:comdata}. To investigate the scale dependency of the scaling relations, we divide the dataset into three populations: GMCs ($R<10 $ pc), MCCs ($10<R<100 $ pc), galaxies ($R>100 $ pc). These sub-divisions contain literature data and our MCCs data. We then plot the velocity dispersion versus radius diagram to describe the first Larson's relation (Figure\,\ref{fig:larson1}), the histogram of virial parameters to describe the second Larson's relation (Figure\,\ref{fig:larsonvirial}), the mass versus radius to present the third Larson's relation (Figure\,\ref{fig:larson3}). Subsequently, we fit linear functions in log-log space to the $\sigma-R$ and $M-R$ to derive their power-laws relations.

As an independent check of the correlation between two parameters, we also calculate the Pearson correlation coefficient $r_{\rm p}$, which is the covariance of the two parameters divided by the product of their standard deviation. An $r_{\rm p} > 0.7$ describes a strong correlated parameter pair, an $ < 0.3 < r_{\rm p} < 0.7$ describes a moderate correlated parameter pair, and an $ r_{\rm p} < 0.3$ describes an uncorrelated parameter pair. The positive sign indicates the positive correlation and vice versa.

\subsection{The $\sigma-R$ relation}
\label{sect:sM}
For the line width-size relation (Figure\,\ref{fig:larson1}), we obtain the following results: 

\begin{flalign}
{\rm GMC:} \,\,\, {\sigma} & = 10^{-0.2}{R}^{
0.4} , r_p = 0.9 \\
{\rm MCC:} \,\,\, {\sigma} & = 10^{-0.6}{R}^{0.7} , r_p = 0.6  \\
{\rm All:} \,\,\, {\sigma} & = 10^{-0.2}{R}^{0.5} , r_p = 0.9  
\end{flalign}

The slope 0.4 of the power law fitted to GMCs is close to $\beta\sim 0.38$ measured by \cite{larson81}.  This is in between the range of the slopes of the pure Kolmogorov turbulent structure ($\sigma\propto R^{0.33}$) and the Burger shock-generated one ($\sigma\propto R^{0.5}$) cases. 
Simulations of cloud formation through turbulent shocks, for examples those of \cite{bonnell06} or \cite{dobbs07}, also produce a $\sigma \propto R^{0.4}$ relation. 
However, the slope 0.7 of the fit to the MCC population is much higher than the GMC's one, even higher than the slope of the Burger shock-generated fractals. 

As for the Pearson correlation coefficient, while the GMC and entire populations produce a good correlation between $\sigma$ and $R$ ($r_{\rm p}=0.9$ and $r_{\rm p}=0.9$, respectively), the MCC population has weaker correlation ($r_{\rm p}=0.6$) and the galaxy population has no correlation ($r_{\rm p}=-0.2$). Similarly, no relation between line-width and size is found for MCCs in nearby galaxies \citep{hughes13,utomo15,rebolledo15}. Together with our MCC data,  we suggest that there is a break at the MCCs scale in addition to an apparent universal $\sigma-R$ relation. However, this break may be artificial due to different measurements such as tracers,
methods, resolution and thresholds for the mass determination were used, or due to the intrinsic different properties of the samples. A more homogeneous investigation is therefore needed to confirm this break.

\subsection{The virial parameter histogram}
 
The second Larson relation describes the tendency of the molecular clouds to reach the virial equilibrium state. We check this hypothesis by examing the virial parameter $\sigma = 5\sigma_{\rm 1D}^{2} R/GM$. A cloud structure is dominated by gravitational energy if $\sigma \le 1$, otherwise external pressure takes control. In Figure~\ref{fig:larsonvirial}, we plot the histogram of virial parameters for different populations. The mean virial parameter of the GMCs is 1.0, of MCCs is 1.8, and of the entire sample is 1.2. Since $\sigma$ of GMCs is closer to 1, the internal gravitational energy is stronger than kinetic energy as oppose to MCCs which are dominated by kinetic energy as their $\sigma$ is larger than 1. In other words, GMCs are closer to be gravitationally bound than MCCs which are super virial and unbound. Similar results were found for MCCs in other galaxies \citep{hughes13}. For the case of our MCCs listed in Table\,\ref{tab:MCCproperties}, their virial parameters are even larger. This picture is also seen in simulations  where most of the large structures are unbound and the small structure are bound \citep{dobbs11,fujimoto14}. The large velocity dispersion may be caused by the fact that MCC is a more dynamic system that can form a compressive environment. Susequently, this results in a higher specific star formation rate or star formation density as we discussed later on.

\subsection{The $M_{\rm gas}-R$ relation}
\label{sect:MR}
The third Larson relation implies that the mass and size obey a power relation $M_{\rm gas}\propto R^\alpha$ with $\alpha\sim2$, or all molecular cloud structures have similar mass surface density, which we examine in Figure~\ref{fig:larson3}. 
The best fits to the GMCs, MCCs, galaxies and entire samples yield slopes of 1.9, 2.2, and 2.0, respectively:

\begin{flalign}
{\rm GMC:} \,\,\, {M_{\rm gas}} & = 10^{2.6}{R}^{1.9} , r_p = 1.0  \\
{\rm MCC:} \,\,\, {M_{\rm gas}} & = 10^{2.1}{R}^{2.2} , r_p = 0.7  \\
{\rm All:} \,\,\, {M_{\rm gas}} & = 10^{2.6}{R}^{2.0}  , r_p = 1.0  
\end{flalign}

All three slopes are very close to $M_{\rm gas}\propto R^2$ although that of MCCs slightly deviates from 2. 
They are in between the value of $\sim~1.6$ for substructures within individual clouds \citep{lombardi10,kauffmann10} and the value of $\sim2.4$ derived for GMCs in the Galactic Plane GRS survey \citep{roman-duval10}. 
The good correlation between mass and size are also present in the large Pearson coefficients: 1.0 for GMCs, 0.7 for MCCs, and 1.0 for the entire population.

The result of this Section is that there is an apparent universal relation between velocity dispersion and size, mass and size of cloud structures across 8 orders of magnitude in size and 13 orders of magnitude in mass. However, there is a break at the MCC scale in the $\sigma-R$ relation and the slopes of individual populations are slightly different. The virial parameters of GMCs and Galaxies are close to 1 while those of MCCs are larger than 1, thus signifies the importance of the kinetic energy contribution in regulating the MCC structures. 

\section{The star formation rates - gas properties relation}
\label{sect:sf}
Similar as in Section\,\ref{sect:larson}, we construct the scaling relations and their best-fitted power law models between the $SFR$ surface density $\Sigma_{\rm SFR}$ and mass surface density $\Sigma_{\rm gas}$, between the SFR and the total mass $M_{\rm gas}$. In addition, we also calculate the Pearson correlation coefficients for different pairs of parameters.  

\subsection{The Schmidt-Kennicutt $\Sigma_{\rm SFR}-\Sigma_{M_{\rm gas}}$ relation}
\label{sect:KS}
\begin{figure*}[hbtp!]
\centering
$\begin{array}{cc}
\hspace{-0.2cm}
\includegraphics[angle=0,width=9cm]{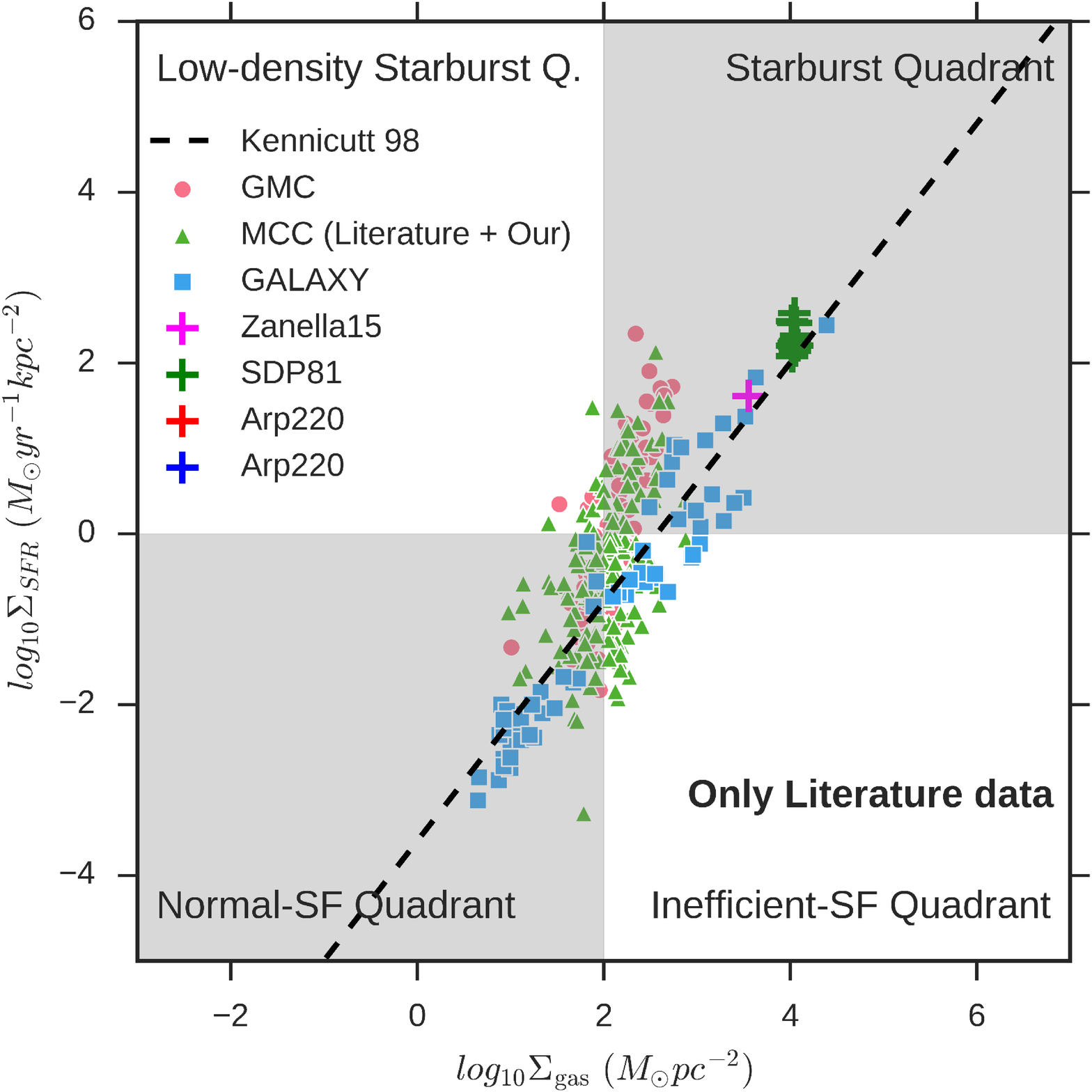} &
\includegraphics[angle=0,width=9cm]{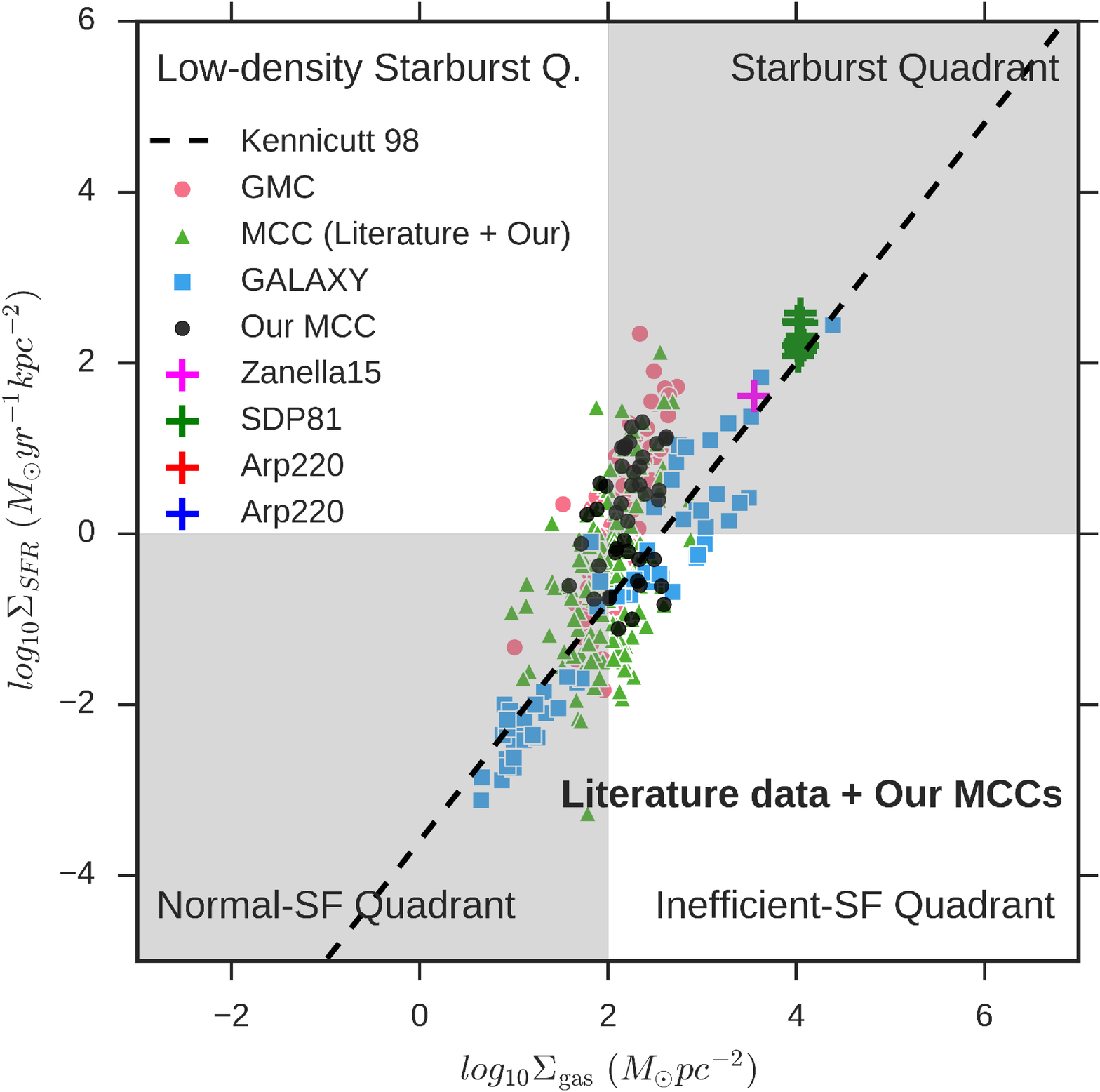} \\ 
\includegraphics[angle=0,width=9cm]{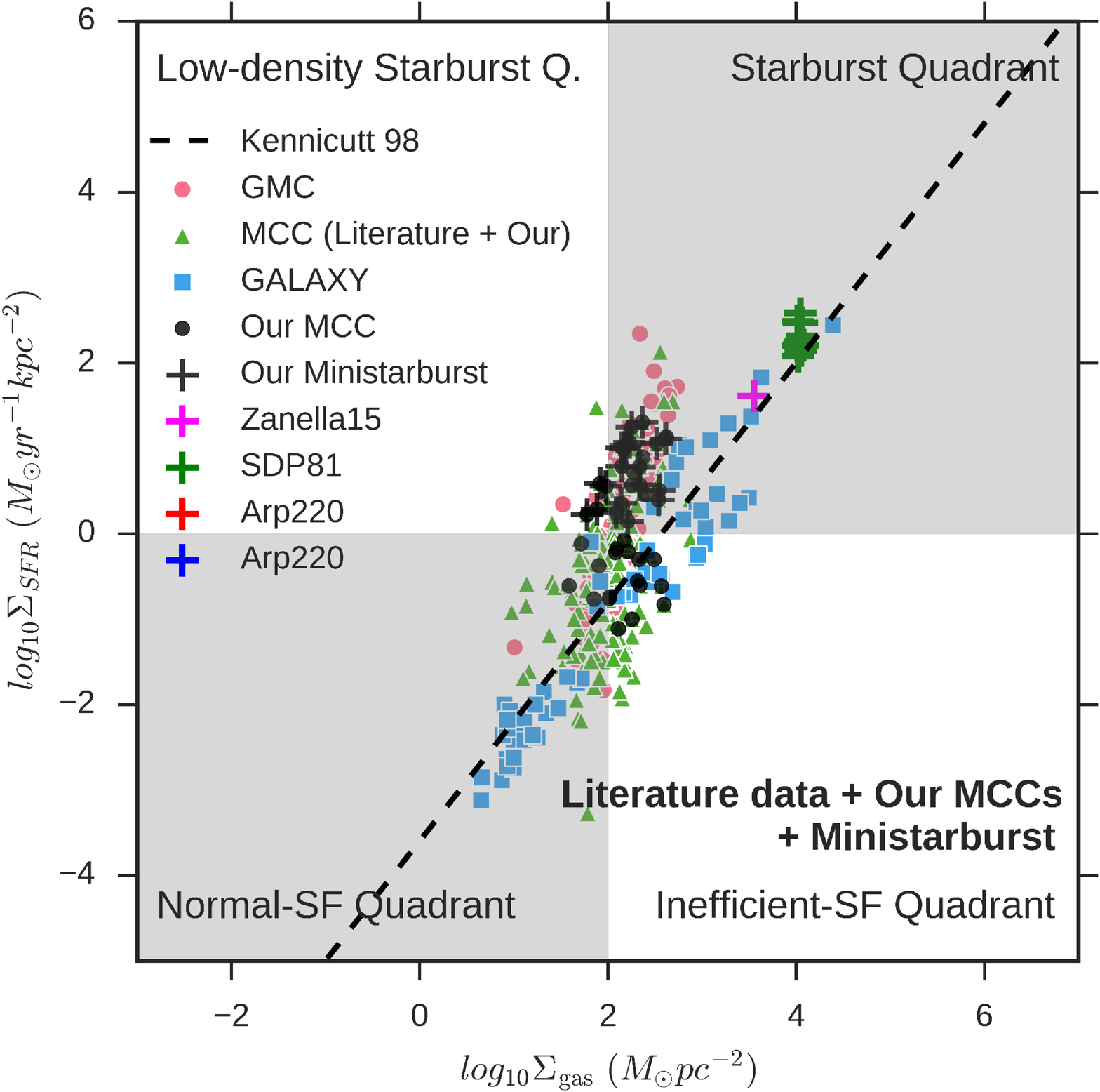} &
\includegraphics[angle=0,width=9cm]{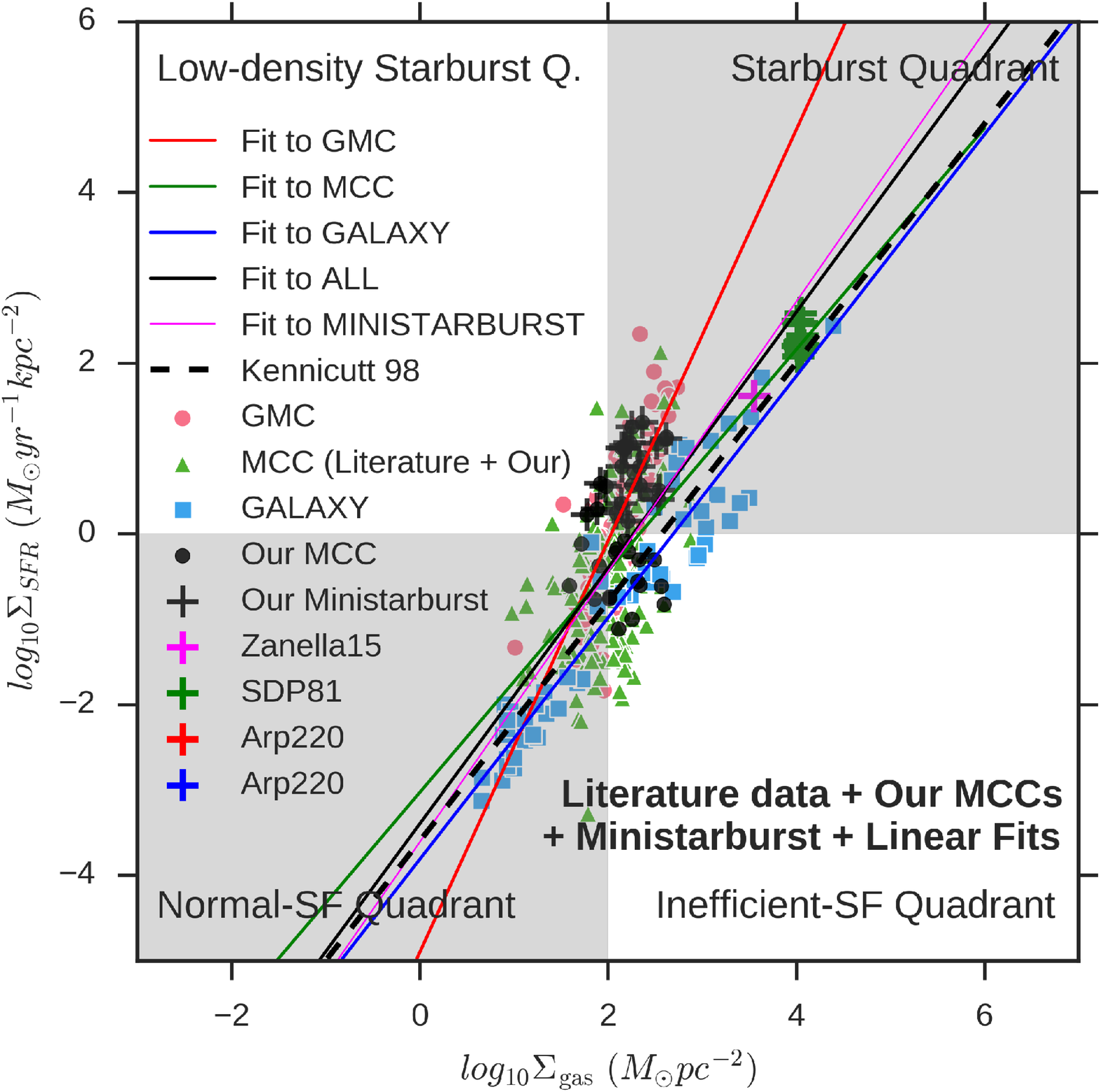} \\ 

\end{array}$

\caption{The Schmidt-Kennicutt relation of $\Sigma_{\rm SFR}$-$\Sigma_{M_{\rm gas}}$ for objects ranging from Milky Way clouds to unresolved galaxies and the best fits. The white and grey regions are divided by the gas surface density of 100~\msun\,pc$^{-2}$ and the SFR density of 1~\msun\,yr$^{-1}$ kpc$^{-2}$. {\bf Top-Left} diagram includes only literature data, {\bf Top-Right} diagram includes literature data and our MCCs from this study, {\bf Bottom-Left} diagram includes literature data, our MCCs and mini-starburst MCCs from this study, and {\bf Top-Right} diagram includes all data and different linear fits as discussed in Equations 10-13 in Section \ref{sect:KS}. We note that the dense cores samples ($R<$ 1 pc) from \cite{maruta10}, \cite{onishi02}, and \cite{shimajiri15} are not included in the SFR analysis due to the lack of SFR measurements.
}
\label{fig:ks1}
\end{figure*}

The $\Sigma_{\rm SFR}-\Sigma_{M_{\rm gas}}$ relation was first constructed by \cite{kennicutt98} for the integrated SFR density $\Sigma_{\rm SFR}$ and the total gas mass density $\Sigma_{M_{\rm gas}}$ of normal star-forming and starburst galaxies. Since then, it was generated as one of the most applicable tool to explain the universal role of gravity in forming stars. Recently, the relation has been extended to GMCs scale \citep{heiderman10,krumholz11} in an endeavour to establish a universal star formation law that connects local to global scales. We re-investigate this relation for our combined dataset by constructing the SFR density and gas mass density relation in Figure\,\ref{fig:ks1} and deriving their best linear fits to the $\Sigma_{\rm SFR}$-$\sigma$ relations in the log-log space as:

\begin{flalign}
{\rm GMC:} \,\,\, {\rm \Sigma_{SFR}} &= 10^{-4.9}{\rm  \Sigma_{M_{\rm gas}}}^{2.4} , r_p = 0.8  \\
{\rm MCC:} \,\,\, {\rm \Sigma_{SFR}} &= 10^{-3.0}{\rm  \Sigma_{M_{\rm gas}}}^{1.3} , r_p = 0.5    \\
{\rm GALAXY:} \,\,\, {\rm \Sigma_{SFR}} &= 10^{-3.8}{\rm  \Sigma_{M_{\rm gas}}}^{1.4} , r_p = 1.0    \\
{\rm ALL:} \,\,\, {\rm \Sigma_{SFR}} &= 10^{-3.4}{\rm  \Sigma_{M_{\rm gas}}}^{1.5}  , r_p = 0.7  
\end{flalign}

The slopes of the $\Sigma_{\rm SFR}$-$\Sigma_{M_{\rm gas}}$ relations are different among different populations. While GMCs have the steepest slope of 2.4, MCCs have the most shallow slope of 1.3, Galaxies have a slope that is closest to the original value of 1.4 derived by \cite{kennicutt98}. 
The steeper slope of the GMC population is consistent with that of the star-forming clump population \citep{heiderman10} and of the resolved individual MCCs \citep{willis15}. 
From a global view of our data, we derive a universal Schmidt-Kennicutt scaling relation with a slope of 1.5 and large scatters in the GMC and MCC populations. The Pearson coefficients show that 
 $\Sigma_{\rm SFR}$ and $\Sigma_{M_{\rm gas}}$ correlate strongly ($r_p =0.9$) in the galaxy population while they correlate least in the MCC population ($r_p =0.5$).

\subsection{The ${\rm SFR}-M_{\rm gas}$ relation}
\label{sect:SFRM}

\begin{figure}[hbtp!]
\centering
$\begin{array}{cc}
\hspace{-0.2cm}
\includegraphics[angle=0,width=9cm]{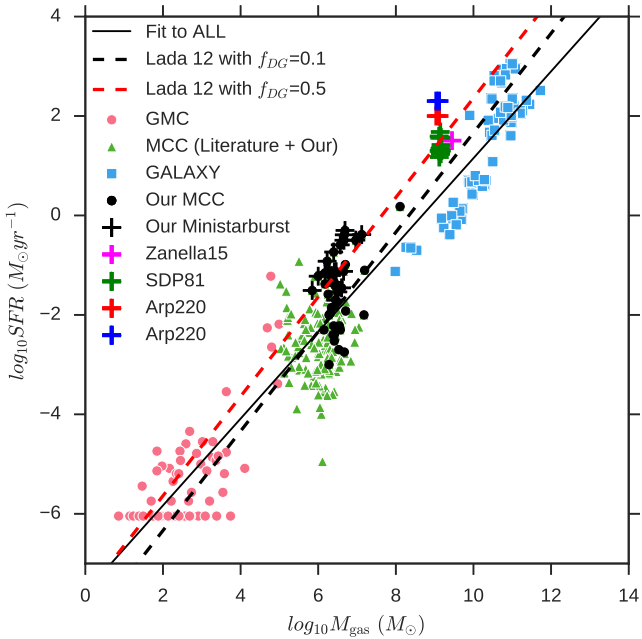} \hspace{-0.2cm}
 \end{array}$
\caption{The SFR$-M_{\rm gas}$ relation for objects ranging from Milky Way clouds to unresolved galaxies and the best fits.  
}
\label{fig:ks2}
\end{figure}

Instead of comparing the surface density quantities of SFR and mass, we examine the integrated SFR and the integrated gas mass relation in Figure~\ref{fig:ks2}. Our literature and new data of MCCs fill the $SFR-M_{\rm gas}$ plane and connect the smallest cloud scale to galaxy scale. 
We plot in Figure~\ref{fig:ks2} the relation between SFR and $M_{\rm gas}$ in the log-log space and fits linear functions to them to obtain the following results for different populations:
\begin{flalign}
{\rm GMC:} \,\,\, {\rm {SFR}} &= 10^{-7.4}{\rm  M_{gas}}^{0.8}  , r_p = 0.7   \\
{\rm MCC:} \,\,\, {\rm {SFR}} &= 10^{-6.2}{\rm  M_{gas}}^{0.6}  , r_p = 0.4   \\
{\rm GALAXY:} \,\,\, {\rm {SFR}} &= 10^{-11.7}{\rm  M_{gas}}^{1.3}  , r_p = 0.9    \\
{\rm ALL:} \,\,\, {\rm {SFR}} &= 10^{-7.6}{\rm  M_{gas}}^{0.9}  , r_p = 1.0  ~~~. 
\end{flalign}

 The slopes of our fits are super-linear for the Galaxy population and sub-linear for all other cases. The slope of the fit to the combined data, is sub-linear with a slope of 0.87, which does not quite agree with the unity slopes derived by \cite{wu05} or \cite{lada10}. Thus to the first order, this universal scaling relation is useful to establish a common relation between SFR and the total gas mass for all star-forming objects. However, at the individual scales, this relation has strong scatter and the linear fits to the data vary strongly. The strong correlation between SFR and $M_{\rm gas}$ of the entire dataset can also be seen in the large Pearson coefficient ($r_p=0.95$). Similar as in the  $\Sigma_{\rm SFR}$-$\Sigma_{\rm gas}$ relation, the MCC has very low Pearson coefficient ($r_p=0.37$), indicating the low correlation between $SFR$ and $M_{\rm gas}$. For example, the total SFR of MCCs vary almost over 4 orders of magnitude. However, they are consistent with the recent numerical simulation of star formation activities in MCCs, these simulation show large SFR spreads in the $SFR$-$M_{\rm gas}$ diagram \citep{howard16}. \cite{lada12} suggested that the spread in $SFR$-$M$ diagram is the effect of the dense gas fraction which we could not address with the current data. {\nlq While we would expect that the slopes of the $SFR$-$M$ relations are super-linear as similar as those of $\Sigma_{\rm SFR}$-$\Sigma_{M_{\rm gas}}$ relations, they are however sub-linear.  The reasons is that the gas surface density also scales with Area as $A\sim\Sigma_{M_{\rm gas}}^{-q}$. Therefore, if there is $SFR\propto M_{\rm gas}^p$, we should get $\Sigma_{\rm SFR}\propto A^{-1}\Sigma_{M_{\rm gas}}^p = A^{p-1}\Sigma_{M_{\rm gas}}^p$. The dependency of gas surface density on area has been investigated by \cite{burkert13}, who showed that $\Sigma_{M_{\rm gas}}$ scales with Area as $A\sim\Sigma_{M_{\rm gas}}^{-3}$ for low-mass star forming regions and $A\sim\Sigma_{M_{\rm gas}}^{-1}$ for massive cores .}

\subsection{The ${\rm SFR}$-$\sigma$ relation}
\label{section:SFR-sigma}

Ideally, if all of the scaling relations in Section~\ref{sect:sM}, \ref{sect:MR}, and \ref{sect:SFRM} hold together, we can deduce a relation between $\Sigma_{\rm SFR}$-$\sigma$ from the other relations. Because $\sigma \propto R^{0.3-0.5}$ and ${\rm  M_{gas}} \propto R^2$, we have the relation ${\rm  M_{gas}} \propto \sigma^{4-7}$. As a consequence of the relation ${\rm {SFR}} \propto {\rm  M_{gas}}^{\left({0.8-1.3}\right)}$, we should have a relation ${\rm {SFR}} \propto {\sigma}^{\left({3.1-8.8}\right)}$. 
However, as shown in Figure \ref{fig:ks3}, we obtain different slopes from fitting power laws to the individual populations and also combined dataset: 

\begin{figure}[hbtp!]
\centering
$\begin{array}{c}
\hspace{-0.2cm}
\hspace{-0.2cm}
\includegraphics[angle=0,width=9cm]{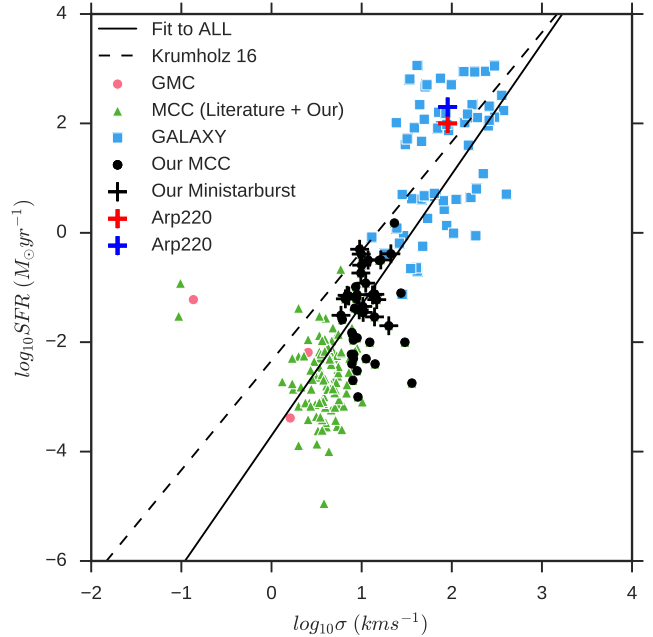} 
 \end{array}$
\caption{The SFR$-\sigma$ (SFR - velocity dispersion) relation for objects ranging from Milky Way clouds to unresolved galaxies and their best fits. There are less GMC data points because there are less GMCs that have both SFR and velocity dispersion measurements.
}
\label{fig:ks3}
\end{figure}

\begin{flalign}
{\rm MCC:} \,\,\, {\rm {SFR}} &= 10^{-2.9}{\rm  \sigma}^{0.9}  , r_p = 0.3   \\
{\rm GALAXY:} \,\,\, {\rm {SFR}} &= 10^{-1.8}{\rm  \sigma}^{1.9}  , r_p = 0.8   \\
{\rm ALL:} \,\,\, {\rm {SFR}} &= 10^{-3.8}{\rm  \sigma}^{2.7}  , r_p = 0.8    
\end{flalign}

The empirical fits show that SFR increases with the velocity dispersion but the slopes are much shallower than the theoretical prediction from other scaling laws. A global fit with a slope of 2.7 fits well to the entire dataset. But fits to GMC and MCC populations are uncertain because their velocity dispersion ranges are less than an order of magnitude. 

\cite{krumholz16} created feedback-driven models and gravity-driven models to explain the SFR$-\sigma$ relations in the galaxy interstellar media and found that the former model produces a steeper relation SFR$-\sigma^2$ while the later model produces a shallower and dense-gas-mass-fraction dependence relation SFR$-f_{\rm DG}\sigma$. These slopes are close to the slope of our galaxy population data, however, they are not applicable to MCCs and GMCs, which seem to be fitted with steeper slopes. 
Similarly at the previous cases, MCCs  have smallest Pearson coefficient ($r_p=0.3$) or their $SFR$ and $\sigma$ are not correlated well.

\section{Discussion \& Conclusion}
\label{sect:discussion}

\subsection{Four quadrants of the Schmidt-Kennicutt diagram}
One of the application of the  $\Sigma_{\rm SFR}-\Sigma_{\rm gas}$ diagram is to distinguish the starburst from the normal galaxies. \cite{daddi10}, for example, argued that the two different regimes of star formation in galaxies are caused by the longer depletion time of normal galaxies than that of the starburst. For the same gas surface density, the starburst galaxies and mini-starburst MCCs have higher SFR density than the normal star forming galaxies and normal star-forming MCCs, therefore lie at a higher location in the Schmidt-Kennicutt diagram. {\nlq  However, the scatter in the $\Sigma_{\rm SFR}-\Sigma_{\rm gas}$ are getting larger as there are better observations, both in galactic and extragalactic scales, which make the recognition of small deviation from the universal law unnoticeable.} Using our combined data, we propose an alternative way to distinguish starburst from normal star-forming objects by applying the $\Sigma_{\rm SFR}$ and $\sigma$ thresholds to the Schmidt-Kennicutt diagram. 

A $\Sigma_{\rm gas}$ threshold of $\sim 100-120~ \msun {\rm pc}^{-2}$ was suggested as the borderline between star-forming and non star-forming clouds or between normal spiral galaxies and starburst galaxies \citep{lada10,heiderman10}. As we can already see in the Schmidt-Kennicutt diagram in Figure~\ref{fig:ks1}, som MCCs or galaxies, and even some GMCs have star formation at  $\Sigma_{\rm gas} < 100~ \msun {\rm pc}^{-2}$. However, this threshold can be safely used as a threshold dividing different star formation modes: isolated versus clustered, normal versus starburst, while the later ones always require more gas with $\Sigma_{\rm gas} > 100-120~ \msun {\rm pc}^{-2}$ \citep{wu05,kennicutt98}. There is exception such as the central molecular zone in the Milky Way, which has high gas mass surface density but low SFR density. 
Additionally, we use the  $\Sigma_{\rm SFR} \sim 1~ \msun {\rm yr}^{-1} {\rm kpc}^{-2}$ as the threshold between starburst and non-starburst objects. This SFR threshold was originally used to distinguish between starburst and normal-star forming galaxies \citep{kennicutt98}.

Putting these two thresholds into the Schmidt-Kennicut diagram, as in Figure~\ref{fig:ks1}, the four quadrants are divided and named clockwise as: low-density starburst quadrant, normal star-forming  quadrant, inefficient-star forming  quadrant, and starburst quadrant. 
Structures in the starburst quadrant have high gas density and high star formation efficiency such as starburst galaxy and ministarburst complex.
Structures in the inefficient star-formation quadrant also have high gas density, however their star formation is impeded. The Central Molecular Zone is a good example of the exception case \citep{immer12}.
The majority of structures in the normal-star formation quadrant are galaxies and they follow well the Schmitt-Kennicutt fit with much less scatter than objects in the starburst quadrant. 
In the low-density starburst quadrant, the low gas density inhibit high SFR density higher than $1~ \msun {\rm yr}^{-1} {\rm kpc}^{-2}$. {\nlq Nevertheless, there are maybe a few candidates that have     $\Sigma_{\rm SFR} > 1~ \msun {\rm yr}^{-1} {\rm kpc}^{-2}$, or they might be the outliers.} 

\subsection{Definition of mini-starburst complexes}
\label{sect:sBness}

The starburst phenomenon was first suggested by observations of the excess star formation activity in galaxy nuclei \citep{arp75,huchra77}. More recently, \cite{elbaz11} proposed that a galaxy experiences a starburst phase if its `current SFR' is twice higher than its SFR averaged over time and used to define its `main sequence SFR'. With a total gas mass of $1\times10^{9}~\msun$ \citep{dame93}, the Milky Way becomes a starburst galaxy only if its SFR is as high as 20 times of the current rate of $\sim 0.7-2~\msun$\,yr$^{-1}$ \citep{robitaille10}. The Milky Way as a whole, is therefore far from experiencing a starburst phase. However, SFR is not uniformly distributed across the Milky Way but excess star formation activity exist in MCC that forms massive star clusters \citep{nguyenluong11b,murray11}.  

{\nlq We define the mini-starburst MCCs as objects having these properties:

\begin{itemize}
 
 \item total gas mass is larger than $10^{6}\,\, \msun$, 
 \item gravitationally unbound,
 \item Star formation rate density is larger than 1\,\,$ \msun {\rm yr}^{-1} {\rm kpc}^{-2}$ or its location is on the starburst quadrant in the Schmidt-Kennicutt diagram  (see Figure~\ref{fig:ks1}).
  \end{itemize} 
 }

From the 44 massive MCCs detected in Section~\ref{sect:souID}, we obtain  21 mini-starburst MCCs. Most of them are the famous mini-starburt MCCs studied extensively in the literatures, for example: 
\begin{itemize}

 \item {\it RCW~106} is
the second brightest MCC in our survey, resides in the Scutum-Centaurus arm and surrounds the bright giant \hii region RCW 106 hosting a rich OB cluster \citep{rodgers1960,nguyenhan15}. 

  \item {\it W43} lies at the meeting point of the Scutum-Centaurus (or Scutum-Crux) arm and the Bar. This mini-starburst is a prototypical example of a mini-starburst MCC \citep{nguyenluong11,carlhoff13}. 
 \item {\it W49} lies on the Perseus arm and hosts ongoing starburst event and forms a very massive star with mass from 100--190\,\msun\,  \citep{galvan-madrid13}.

 \item {\it Cygnus X} is one of the most massive mini-starburst MCC and is located in the Cygnus arm \citep{schneider06}. 

 \item {\it W51} is near the tangent point of the Sagittarius arm, has a high dense gas fraction and host massive star formation events  \citep{ginsburg15}.

 \end{itemize}

We compare the Galactic mini-starburst with the extragalactic mini-starburst which are resolved to a comparable scale.
First, we use data from Arp 220, a relatively nearby (d$\sim 75$~Mpc) ultraluminous infrared galaxy \citep{soifer84}, which  contains two nuclei that are powered by extreme starburst activity. \cite{sakamoto08,wilson14,scoville15} used ALMA to resolve the Arp 220 down to a spatial scale of $\sim 100~$pc and obtain gas mass densities of $5.4\times10^4$ and $\rm 14\times10^4 ~ \msun pc^{-2}$ for the Eastern and Western nucleus. Taking into account the SFR of $\rm 100-200 ~\msun yr^{-1}$ \citep{scoville15}, we obtain a SFR density  $\rm 10{^4}-10{^{4.5}} ~ \msun yr^{-1} kpc^{-2}$, the highest SFR density at the MCC scale.   
Second, we use data of 14 starburst MCC clumps in SDP.81 galaxy derived from ALMA observations. 
SDP.81 is one of the bright galaxy at a redshift $z = 3.042$ (or a luminosity distance of $\sim 25\times10^{3}$ ~ Mpc) and is gravitationally lensed by a foreground galaxy at z = 0.2999, therefore it allows us to resolve the gas properties down the scale of $\sim 200$ \,pc by ALMA \citep{hatsukade15}. 
Finally, we also include in our comparison the data from the first direct measurement of a MCC  clump at z = 1.987 down to the scale of $\sim 500$\,pc \citep{zanella15}.
Although they are larger than our mini-starburst, we still compare them with our data keeping in mind that their $\sigma$ and $\Sigma_{\rm SFR}$ can be higher if we resolve them at a smaller scale. 

The SFR and SFR density of the extragalactic mini-starburst are ten to hundred times higher than the Galactic counterparts that they are located in the upper part of the starburst quadrant and the upper part of the SFR$-M_{\rm gas}$ diagram. Highly-compressed gas in these extragalactic mini-starbursts may be the origin of their  active star formation activities.

\subsection{Dynamical Evolution of mini-starburst MCCs}

We advocate that the high $\Sigma_{\rm SFR}$ and high $\sigma$ of a mini-starburst MCC is caused by dynamical processes happen during the MCCs evolution. These processes are supported by externally induced pressure such as shocks, galactic disk gravitational instability or colliding flow
\citep{vazquez-semadeni96,bonnell13}. Therefore, mini-starburst MCCs are often found at highly dynamics regions such as the overlapping regions  in Antennae galaxy \citep{herrera12,fukui14} or at the end of the Galactic Bar \citep{nguyenluong11}. Simulation of molecular cloud evolution in a galaxy also agrees with this view by showing that MCCs are concentrated mostly in the Bar or spiral arms, and have high SFR \citep{fujimoto14}.

\begin{figure}[hbtp!]

\centering
$\begin{array}{c}
\hspace{-0.2cm}
\hspace{-0.2cm}
\includegraphics[angle=0,width=9cm]{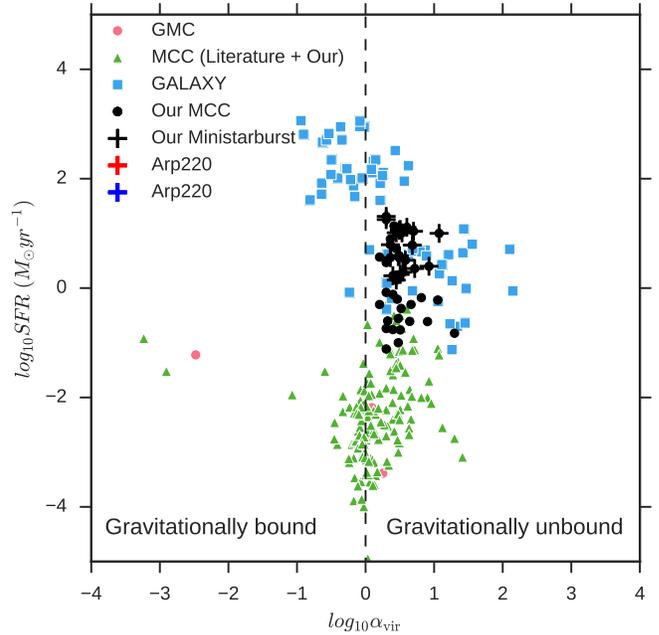} 
 \end{array}$
\caption{The SFR$-\alpha_{vir}$ (SFR - virial parameter) relation for objects ranging from Milky Way clouds to unresolved galaxies. It shows that there is no particular relation between SFR and $\sigma$ but the mini-starburst MCCs are mostly gravitationally unbound. There are less GMC data points because there are less GMCs that have both SFR and velocity dispersion measurements.
}
\label{fig:ks4}
\end{figure}

In all cases, continuous gas flows agglomerate clouds, compress material and develop active star formation sites \citep[e.g.,][]{koyama00,bergin04}, which  explains why mini-starburst MCCs have high SFR (see Sect.~\ref{sect:sBness}). This framework is also called cloud-cloud collision, advocated as the main formation mechanism of massive star and stellar cluster \citep{inoue13}. For example, in W43, one of the most active massive star-forming regions, both large-scale and small-scale gas flows were observed as a mean of forming dense gas and massive stars \citep{motte14,nguyenluong13,louvet14}.

In addition, a super-linear relation between $SFR-\sigma$ (Figure~\ref{fig:ks3}) indicating that SFR increases with the turbulence or compression degrees of the gas supports the dynamical view of MCC evolution. The steep slope of the $SFR-\sigma$ relation, especially that of MCC, disagrees with the slope produced by the gravity-driven models or feedback-driven model
\citep{krumholz16}. Therefore, cloud compression plays a stronger role in controlling the structure and the star formation activity, beyond gravity and feedback. As a consequence, the majority of MCCs are gravitational unbound and form stars efficiently as shown in Figure~\ref{fig:ks4}, especially th mini-starburst MCCs.

\section{conclusions}
\label{sect:conclusions}

We investigated the connection between the local and global star formation by comparing the mass, size, line width, and star formation rate of cloud structures across 8 orders of magnitude in size and 13 orders of magnitude in mass. 
Our focus is on molecular cloud complexes (MCCs), which have radii of $\sim$50--70~pc and masses $>10^6~\msun$.
 We use the $^{12}$CO 1--0 CfA survey to identify and characterize a sample of 44 MCCs in the Milky Way (see Table~\ref{tab:MCCproperties}). This sample is complete up to a distance of 6~kpc from the Sun. Their distribution follows the spiral arms, especially the Scutum-Centaurus and Sagittarius arms (see Figures~\ref{fig:MScanGP}-\ref{fig:MSfaceon}). 

Together with data from the literature, we reproduced the scaling relations and the star formation laws: $\sigma-R$, $M_{\rm gas}-R$, $\Sigma_{\rm SFR}-\Sigma_{M_{\rm gas}}$, ${\rm SFR}-M_{\rm gas}$, and  ${\rm SFR}-\sigma$. 
Apart from being apparently universal, the slopes and the coefficients are different for individual scales: GMC, MCC, and galaxy. Second, there is a break at the MCC scale in the $\sigma-R$ relation and a break between the starburst objects such as mini-starburst, star-forming clumps from the normal star-forming objects in the SFR-$M_{\rm gas}$ and $\Sigma_{\rm SFR}$-$\Sigma_{M_{\rm gas}}$ relations. 

These breaks enable us using the Schmidt-Kennicutt diagram to distinguish the starburst from the normal star-forming objects by using the $\Sigma_{M_{\rm gas}}$ threshold of 100\,\,\msun pc$^{-2}$ and the $\Sigma_{\rm SFR}$ threshold of 1\,\,\msun yr$^{-1}$ kpc$^{-2}$. These two thresholds divide the  $\Sigma_{\rm SFR}-\Sigma_{M_{\rm gas}}$ diagram into four quadrants: Q1 as low-density starburst quadrant, Q2 as normal star-forming  quadrant, Q3 as inefficient-star forming  quadrant, and Q4 as starburst quadrant. 
{\nlq Mini-starburt MCC are gravitationally unbound MCCs that have enhanced SFR density that is larger than $1\,\msun$ yr$^{-1}$ kpc$^{-2}$.}

We propose that mini-starburst MCC is formed through a dynamical process, which enhance the compression of clouds and induce intense star formation as bursts and eventually form young massive star cluster. 
Because of the dynamical evolution, gravitational boundedness does not play a significant role in characterizing the star formation activity of mini-starburst MCCs.
Therefore, there is no particular relation between SFR and the virial parameter (see Figures ~\ref{fig:ks3}-\ref{fig:ks4}).

\begin{acknowledgements}
QNL acknowledges the financial support from the
East Asian Core Observatories Association
(EACOA) through the EACOA fellow program and the support from Canadian Institute for Theoretical Astrophysics during his visit at CITA. N.S. acknowledges support through the DFG project numbers  0s 177/2-1 and 177/2-2, and central funds of the DFG-priority program ISM-SPP. We thank  the anonymous referee and Neal J. Evans for giving constructive comments which improve the quality of the paper.
\end{acknowledgements}

\bibliographystyle{apj}  \bibliography{quangreference}

\end{document}